\documentstyle[12pt]{article}
\begin{document}
\begin{center}
{\bf{ Electronic Structure of a Chemisorbed Layer at Electrochemical \\ Interface
:  Copper Layer on Gold Electrode  \\}}

\vskip 1.2cm

{\bf A. K. Mishra } \\

\vspace{.5cm}

{\it Institute of Mathematical Sciences \\
C.I.T.Campus, Madras-600113, India.} \\
\end{center}

\vspace{1.5cm}

\baselineskip=20pt
\centerline{\bf ABSTRACT}

\vskip .5cm

\noindent Though the importance of chemisorption at the electrochemical interface is well
recognized, an electronic level description for the same still remains in a
nascent stage. The present work specifically addresses this problem. An
appropriate model Hamiltonian based formalism is proposed for a random adsorbate
layer with arbitrary coverage and the ensuing two-dimensional band formation by metallic
adsorbates in the monolayer regime. The coherent potential approximation is
employed to handle the randomness. The adsorbate self-energy is evaluated
explicitly using the density of states for the substrate band. This takes us
beyond the conventional wide-band approximation and removes the logarithmic
divergence associated with the binding energy calculations. The formalism
is applied to the electrosorption of copper ion on gold electrode, and the coverage dependence
of adsorbate charge, binding energy, and adsorbate density of states are determined.
The analysis predicts a unique charge configuration of copper adsorbate, having a 
net positive charge, in the high-coverage regime, and multiple charge states
when the coverage is low. Though one of the charge configurations of copper is nearly
neutral at small coverage, its  positive charge state is the most stable in the entire 
coverage range. The transition from the relative neutral state of copper at low coverage to a 
positive charge configuration  occurs sharply at the intermediate-coverage region.
This transition is caused due to the progressive desolvation of copper 
adion with increasing coverage. The energy calculations show that copper s-orbital
bonding contributes maximally towards the binding energy at the metal-vacuum interface,
whereas it is the copper d-orbital bonding which makes chemisorption feasible
at the gold electrode. Finally, our numerical results are compared with the
relevant experimental studies.

\newpage

\section{Introduction}

~~~~Chemisorbed species significantly modify the physico-chemical nature of an
electrochemical interface [1-3]. In spite of their importance in the areas of heterogeneous
electron-transfer reactions, catalysis, chemically modified electrodes, and 
double layer studies, an electronic level description of the electrosorption phenomenon
is still elusive. 
However, the recent progress concerning  in situ structure-sensitive techniques
has made it possible to obtain the  structural details of an adlayer in an
electrochemical environment. In the present paper, we investigate the electronic 
states of a random adsorbate layer.   
To understand the similarities, as well as the critical
differences between the chemisorption at the metal-vacuum and
electrochemical interface, is the other motivation for undertaking this study.


In our studies, the adsorbate coverage factor $\theta$ is allowed to take any
arbitrary value in the range $(0,1)$. Thus the formalism remains valid all the way
up to a monolayer regime, starting with a single adsorbate case. The analysis
ultimately leads to the quantitative estimates of (i) average ``occupation
probability'' and the ``partial charge-transfer coefficient'' of adsorbate, (ii) its
density of states, and  (iii) binding energy. The formalism also enables one to
study how these quantities are affected due to coverage variation.


Electrochemisorbed copper on gold electrode represents a model experimental system,
studied extensively through various structure-sensitive techniques [4-7]. Besides,
this is also typical of the underpotential deposition phenomenon [8-10]. 
Double layer studies coupled with the thermodynamic analysis suggest that a copper 
adsorbed on gold electrode exists in a neutral state when the coverage is low [11,12].
On the other hand, Tadjeddine et al. have shown through in situ  X-ray absorption
spectroscopy of copper layer on gold(111) electrode surface that in the monolayer regime,
the charge state of Cu adsorbate is close to $Cu^{+}$ [6]. These studies show a large
variation in copper adatom charge with increasing coverage. It will be pertinent
here to restate that the electronic structure of a metallic adsorbate  differs
significantly in the low and higher coverage domains. In the former case, the electrons
are essentially localized on the adatom whereas with increasing coverage, they tend
to get delocalized over a large spatial region. These extended electronic states
ultimately form a two-dimensional band in the monolayer regime. We have applied
our general formalism to Cu/Au system. Our theoretical results are in conformity
with the Tadjeddine et al. experimental observation of a positively charged 
copper adsorbate in the high coverage regime. For the low-coverage domain, analysis
shows the existence of three different charge states of a copper adsorbate.
Though one of these states corresponds to a nearly neutral configuration of the copper,
energetic considerations make this state a metastable one. Also the transition from
this  neutral to a positive charge configuration is not smooth, but occurs discontinuously in
the intermediate-coverage region. We show in sections 7 and 8 that the progressive desolvation
of a metallic adsorbate with increasing coverage is the main reason behind
this feature. The transition from multiple charge states of copper when coverage
$\theta$ is small to a single charge state for intermediate value of $\theta$
is specific to electrochemical interface. 
This indicates that the chemisorption at the electrochemical and 
metal-vacuum interface can be qualitatively different, even when  the adsorbate
and substrate  remain the same. In the context of multiple charge states of copper, 
we also note  that when the net charge on an adsorbate
at the electrochemical interface gets reduced, its solvation energy decreases. This in turn 
shifts adsorbate's energy toward a more positive value.


The present  analysis employs, as the model Hamiltonian, a generalized version of the Anderson-Newns
Hamiltonian [13-16], which has been studied extensively in the context of
chemisorption at a metal-vacuum  interface. The generalization considered here
refers to  (i) modeling additional interactions which an adsorbate experiences at an
electrochemical interface, (ii) considering  a collection of adspecies along 
with their mutual interactions, instead of a lone adsorbate, and (iii) allowing a
randomness in the occupancy status of adsorption sites. 
Obviously the formalism considered here is also valid for the adsorption at the 
metal-vacuum  interface in the appropriate limiting conditions.


Kornyshev and Schmickler (KS) have earlier considered the coverage dependence of 
the adsorbate charge using a generalization of Anderson-Newns Hamiltonian, and
have applied their formalism to study the adsorption of Cs and I at a mercury
electrode [17]. The coverage dependence in their analysis arises due to nonlocal
electrostatic interactions between adions.

The present formalism  differs qualitatively from Kornyshev and Schmickler approach,
both at the levels of Hamiltonian (cf. section 2) and  analysis.


In the KS formalism, the other adsorbates essentially introduce a static shift 
in the energy level of an adatom. The quantum-mechanical aspect of the adlayer problem
is still handled within a ``single-adsorbate'' model. 
In contrast, the present formalism takes into account the band structure of the 
metallic adlayer. Next, the KS approach is based on the  wide-band approximation 
which has been extensively employed  to study the chemisorption at the metal-vacuum (M-V)
and electrode-electrolyte (E-E) interface. Herein, a part of adsorbate's 
self-energy arising due to the metal-adsorbate hybridization is approximated as an
energy independent quantity $i\Delta$. This approximation leads to a logarithmic
divergence in the energy calculation and necessitates an introduction of a lower cutoff
parameter in order to arrive at a finite result [18].
Within the wide-band approximation, the adsorbate Green's function has a single pole
structure (Lorentzian density of states) when the intraadsorbate Coulomb repulsion 
is treated using the Hartree-Fock approximation.

We have gone beyond the wide-band approximation in the present formalism.
The relevant self-energy arising due to the electronic  
coupling between metal and adsorbate states is explicitly evaluated employing model 
density of states for the substrate. This removes the divergence in the energy
calculation and, depending on the system parameters, allows the adsorbate Green's function  
to admit more than one pole even within the Hartree-Fock limit. 
The self-energy now depends on the energy variable, and  the adsorbate density of states
may exhibit non-Lorentzian structure.

In order to substantiate as well as compare our results 
with the wide-band limit based formalism, we have  suitably 
extended the latter to the context of random adsorbate layer (sections 6 and 8).



In the next  section, a   model Hamiltonian  describing the 
electrochemisorbed layer is constructed. The coherent potential approximation  (CPA)
scheme for the random adsorbate layer is considered in  section 3. The  adsorbate
density of states and other relevant quantities are evaluated in the next
two sections.  Analytic  expressions  suitable for the numerical computations are
derived  in section 6 and the application of the formalism to the copper adsorption on gold
electrode is considered in section 7. Section 8 is devoted to discussion, followed by  section 9
on summary and concluding remarks. 


\section{The Model Hamiltonian}


~~~~Chemisorbates forming a random layer over an electrode surface have extensive 
electronic coupling with the substrate band states as well as with other adspecies.
In addition, they interact with the orientational, vibrational, and electronic
polarization modes of a polar liquid in the electrolyte phase. This leads to the
solvation energy of the adsorbate. Similarly, the coupling between the adsorbate and surface
plasmons in the substrate is responsible for the image energy contributions. The various
solvent polarizations and surface plasmon modes are described within the harmonic
approximation, and their coupling with the adsorbates is modeled as diagonal
fermion-boson terms . Taking into account all the above
interactions along with the intraadsorbate Coulombic repulsion, a Hamiltonian for
the electrochemisorbed system is  written as [16,17]
$$ \begin{array}{lcl}
H & = & \sum_{k, \sigma} \epsilon_k n_{k\sigma} + \sum_{i,\sigma} \hat \epsilon_{i\sigma}
(\{b_\nu + b^{\dagger}_{\nu}\}, U n_{i\bar{\sigma}} ) n_{i\sigma} \\

\\
& & + \  \sum^4_{\nu =1} \omega_\nu b^{\dagger}_\nu b_\nu \\ 
\\
& & + \sum_{k,i,\sigma} \{v_{ik} c_{i\sigma}^{\dagger} c_{k\sigma} + h.c.\} + \sum_{i \ne j
, \sigma }v_{ij} c_{i\sigma}^{\dagger} c_{j\sigma} \\ 
\\
& & - \sum^4_{\{i\},\nu=1} \{\lambda_{ic\nu} (b_\nu + b^{\dagger}_\nu) 
+ U n_{i\sigma} n_{i\bar{\sigma}}  \}
\end{array}\eqno(2.1)
$$

The first two terms describe the electronic states of electrode and the
adsorbate layer. $ \epsilon_k$ and $\hat \epsilon_{i\sigma} n_{i\sigma}$ are the energy of $k$th
electronic level in the substrate and the energy operator for the site $i$ in the two-dimensional adsorbate lattice.
$\hat \epsilon_{i\sigma}$ includes the contributions to the  adsorbate orbital energy due to its 
coupling to various boson branches, and 
from the intraadsorbate coulomb repulsion $U$.
$c^{\dagger}_\ell(c_\ell)~$ and $n_\ell$ are creation (annihilation) and number operators for
the quantum state $\ell$ . The set $(\{k\sigma\}, \{(i\sigma\})$ is assumed to form a
complete orthonormal set. The third term corresponds to the three polarization
branches in the solvent medium $(\nu = 1,~2,~3)$ and the surface plasmon in the
substrate $(\nu = 4)$. The next two entries describe the respective hopping terms
between the sites on adsorbate layer and substrate band states, and between the adsorbates 
in the two-dimensional lattice. The coupling between the adsorbate core charge and the
boson modes is represented  by the last term in eq 2.1. The $U$ term here removes the
double counting of the intraadsorbate repulsion. $\{i\}$ implies that the 
summation is over the sites occupied by the adsorbates.

\vskip .5cm

At an electrochemical interface, the solvent layer adjacent to electrode is usually
described, to a first approximation, as a structureless  medium of
dielectric constant 5.0 [19]. There is no charge transfer  between the
dielectric medium and the electrode. It only provides  the background in
which the adsorbate lattice is embedded. The operator  $\hat \epsilon_{i\sigma}$, when
a site $i$ is occupied by an adsorbate, is [20]
$$ 
\hat \epsilon_{i\sigma}(\{b_{\nu} + b^{\dagger}_{\nu}\}, U n_{i\bar{\sigma}}) \ = \ E_{a} + \  \sum^4_{\nu=1}
\lambda_{ia\nu} (b_\nu + b^{\dagger}_\nu) + Un_{i\bar{\sigma}} \eqno(2.2)
$$
where $E_{a}$ is the adsorbate's orbital energy in the vacuum. 


In case site $i$ is unoccupied by an adsorbate, the expectation value of
$\hat \epsilon_{i\sigma}$ is given by the relation 
$$
 <\hat \epsilon_{i\sigma}> \longrightarrow \infty \eqno(2.3)
$$
This ensures no charge sharing between the electrode and  vacant sites.

\vskip .5cm

$\hat \epsilon_{i\sigma}$ is therefore a random entity, giving rise to a diagonal randomness in the
problem. All other parameters, both diagonal and non-diagonal, are taken to be
deterministic. They do not depend on the occupancy status of the sites.

\vskip .5cm

It is pertinent here to consider how a system described through the above  Hamiltonian differs 
from the system studied by Krastov and Mal'shukov (KM) for the structural phase transition 
in an adsorbate layer [21]. This  clarification is also needed  because the  approach 
used by KS to study the coverage dependence of the 
adsorbate charge at an electrochemical interface [17]  has a similarity with the KM formalism. 
Therefore, we mention the following basic
differences between the two approaches :

\vskip 0.2 cm

(i) We consider a random distribution of adsorbate on the 2D
lattice. No such randomness is inherent in KM formalism. We treat the
randomness using the CPA approach. Being an effective  
medium theory, CPA averages out all the inhomogeneities in the system and
provides us with a description in which all the adsorption sites are equivalent;
though the effective self-consistent site energy is now a function of energy variable
(cf. section 3).

\vskip 0.2 cm

(ii) We model the direct interaction between adsorbate through the
hopping term $ \sum_{i \ne j, \sigma} v_{ij} c^{\dagger}_{i \sigma} c_{j
\sigma}$. This ``one-body'' interaction  leads to the band structure of 
2D adlayer (cf. section 3).


In contrast, the interadsorbate direct interaction is modeled
in the  KM analysis through a ``two-body'' Coulomb term $ (1/2) \sum_{i\ne j} U \hat n_i \hat n_j$
describing the electrostatic interactions between the adions. The 
same is true for the KS studies on the coverage dependence of the 
adsorbate charge.

\vskip 0.2 cm

(iii) In the KM formalism, the shift in adsorbate orbital energy
due to a boson mode is $ \lambda^2 / \omega_o <n> $ for all coverages.
The coverage dependence of this shift comes only through the coverage
dependence of the average charge $<n>$. To take into account the progressive desolvation
of the adspecies 
with increasing coverage, the ensuing shift due to solvent polarization modes is scaled
by a factor $( 1 - \theta^2)$ in our formalism (cf. section 7).

\vskip 0.2 cm
(iv) The  KM  analysis employs ``spin-less'' approximation wherein
the intraadsorbate coulomb repulsion $U$ is taken to be the  largest energy parameter in 
the system. Thus, only one electron is allowed in a valence orbital. In Hamiltonian (2.1), 
no such approximation has been used.  

\vskip 0.2 cm

(v) Both the  KM  and KS analysis  are  based on the wide-band approximation.

\vskip 0.3 cm

The above comparison shows that present formalism is qualitatively
different from the KM and KS  approach for the coverage-dependent properties 
of an adlayer. However, a suitable extension of the wide-band limit based approach
for random adlayer would enable us to make a comparison between the present
and the KM, KS formalism, at least when the coverage is small. We refer to
sections 6 and 8 for further details in this regard.

\vskip .5cm

To summarize, the chemisorbed species are considered here to be distributed randomly over
various adsorption sites on the substrate lattice. These sites form a 2D lattice in a
medium of effective dielectric constant 5.0. We take this 2D
lattice to be commensurate with the underlying substrate surface. This implies that
the adspecies occupy ``on-top'' positions. At coverages less than unity, the sites
unoccupied by adsorbates constitute the second component of the adlayer, the first
being the occupied sites. At complete coverage $(\theta = 1)$ the adsorbates form $(1
\times 1)$ ordered structure. The various  system components and their 
interactions are  described by the Hamiltonian (2.1).

\vskip .5cm

\subsection{An Effective Hamiltonian.}
The Hamiltonian (2.1)  presents a complex problem, the reason
being its many-body nature and the inherent randomness in the system. The many-body 
aspect follows from the intraadsorbate coulomb repulsion and the fermion-boson interactions.
We treat the former within the Hartree-Fock approximation, which in effect raises  the
$E_a$ by $U < n_{a\bar{\sigma}} >$. In an earlier communication [16], we had
analyzed the fermion-boson coupling in a detailed manner through superoperator
technique. Therein, it had been  shown that for low frequency boson modes, a mean
field decoupling
$$
\phi n_{a\sigma}~ \approx ~  < \phi > n_{a\sigma} + \phi < n_{a\sigma} >   \eqno(2.4)
$$
suffices $(\phi = b+b^{\dagger})$. Consequently, $E_a$ gets  further shifted by
$\sum_{\nu=1,2}  \lambda_{i\nu} < \phi_{\nu} >$ where $\nu = 1,~2$ respectively refers to
orientational and vibrational polarization modes of the solvent. For high frequency
modes, it becomes  necessary to go beyond the mean field approximation. The resulting shift in
$E_a$ is now $\sum_{\nu=3,4} [2\lambda_{ic\nu} \lambda_{i\nu}/ \omega_{\nu}  - 
\lambda_{i\nu}^2/ \omega_{\nu}]$, where indices 3 and 4 correspond to electronic 
polarization branch in the solvent and surface plasmon oscillations in the substrate. 

\vskip .5cm

With these simplifications, and after substituting the value of $< \phi_{\nu} >$ [16], the
Hamiltonian (2.1) gets replaced by an effective Hamiltonian: 

$$ \begin{array}{lcl}
H & = & \sum_{k, \sigma} \epsilon_k  n_{k\sigma} + \sum_{i,\sigma} \epsilon_{i\sigma}
n_{i\sigma} + \sum_{i,k,\sigma} \{v_{ik} c^{\dagger}_{i\sigma} c_{k\sigma} + h.c.\} \\
\\
& + & \sum_{i \ne j, \sigma} v_{ij} c^{\dagger}_{i\sigma} c_{j\sigma} \\
\\
& + & \  \sum^4_{\nu=1}\omega_\nu b^{\dagger}_\nu b_\nu  \\
\\
& + &\  \sum_{\{i\}, \nu = 1,2} ( \lambda_{i\nu} ( \sum_{\sigma} <n_{i\sigma}> -{\lambda_{ic\nu}})
) \phi_\nu \\
\\
& - & \sum_{\{i\}} (U - 2 \sum_{\nu = 3,4} \frac{\lambda^2_{i\nu}}{\omega_\nu}) <n_{i\sigma}>
<n_{i{\bar\sigma}}>
 \\
\\
& - & \sum_{\{i\},\sigma, \nu =1,2} \frac  {\lambda_{i\nu}^2}{\omega_{\nu}} \left (
\frac{2\lambda_{ic\nu}}{\lambda_{i\nu}} - \sum_\sigma <n_{i\sigma}>\right ) < n_{i\sigma}>
\\
\\
& - & \sum_{\{i\}, \nu = 3,4} \frac{\lambda^2_{ic\nu}}{\omega_\nu}
\end{array} \eqno(2.5)
$$
where
$$ \begin{array}{lcl}
\epsilon_{i\sigma} \equiv \epsilon_{a\sigma} & = & E_a + \sum_{\nu = 1,2} \frac{2\lambda_{i\nu}^2}{\omega_{\nu}}
\left( \frac{\lambda_{ic\nu}}{\lambda_{i\nu}} - <n_{i\sigma}> \right) \\
\\
& + & \sum_{\nu = 3,4} \left( \frac{2\lambda_{ic \nu} \lambda_{i\nu}}{\omega_\nu} -
\frac{\lambda_{i\nu}^2}{\omega_\nu} \right) \\
\\
& + & \left (U - 2 \sum^4_{ \nu=1} \frac{\lambda_{i\nu}^2}{\omega_{\nu}}\right ) < n_{i\bar\sigma} >
\end{array} \eqno(2.6)
$$
when a site $i$ is occupied by an adsorbate; and as noted earlier, $\epsilon_{i\sigma}$ tends to infinity 
otherwise. $\{i\}$ implies a summation over occupied sites only. The second and third
terms on the right side of (2.6) describe the shifts in the adsorbate orbital
energy due to slow and fast bosons modes respectively. The fourth term takes into
account the renormalization of $U$ caused by fermion-boson interactions. While
evaluating the shift in $\epsilon_{i\sigma}$ due to bosonic coupling, the boson-mediated
interactions between different sites are neglected. Next, since the adsorption of a
single species is considered, we replace $\lambda_{i\nu}$ by $\lambda_{a\nu}$ and
$\lambda_{ic\nu}$ by $\lambda_{c\nu}$. 

\section {Green's Function : A CPA approach}

~~~~While constructing the effective Hamiltonian (2.5) from (2.1), two important simplifications
are made. (i) A decoupling between the fermionic and bosonic components is achieved.
(ii) The system is now described by  a one-body Hamiltonian.  But the problem associated with 
the random nature of $\{\epsilon_{i\sigma}\}$ still remains. We tackle it using the concept of
coherent potential approximation  [20,22,23]. Herein, the random adsorbate layer
is replaced by an effective, non-random, 2D medium. This is achieved by
replacing the random variable $\epsilon_{i\sigma}$ by a deterministic energy parameter $k_{\sigma}(\epsilon)$
which is same for all the sites. $k_{\sigma}(\epsilon)$ depends on the energy variable $\epsilon$. 
It has to be evaluated self-consistently using  the condition that the
configuration averaged $t$-matrix for an arbitrary site $i$, i.e. $<t_i >_c$ is
zero. $<...>_c$ denotes the configuration average. The physical reasoning behind this
step is linked to the fact that random potentials at different sites cause a
scattering of incident particles. In an effective medium, potential is the same at all
the sites and hence there is no scattering; thus $<t_i>_c = 0$.

The fermion Green's function (GF) matrix elements corresponding to sites on adsorbate layer
satisfy the relation [20,23]
$$
G_{ij} \ = \ G^o_{ij} \delta_{ij} + \sum_\ell G^o_{ii} W_{i\ell} G_{\ell j} \eqno(3.1)
$$
where 
$$
G^{o\sigma}_{ij} \ = \ < c_{i\sigma} \frac{1}{\epsilon-\sum_{i\sigma} \epsilon_{i\sigma} n_{i\sigma}}
c_{j\sigma}^{\dagger} >  \eqno(3.2)
$$
$<...>$ denotes the quantum mechanical average. The unperturbed GF is :
$$
 \begin{array}{lcl}
G^{o\sigma}_{ij} & = & \frac{\displaystyle 1}{ \displaystyle \epsilon - \epsilon_{a\sigma}} ~~~~ \rm {if}~i = j~~ 
\rm{and~site~i~is} \\
& & ~~~\rm{ ~~~~~~~occupied ~ by ~ an ~adsorbate} \\
\\
& = &  0 ~~~~~~~~~\rm{otherwise} \end{array} \eqno(3.3)
$$
$$
W_{i\ell} = v_{i\ell} + \sum_k \frac{v_{ik} v_{k\ell}}{\epsilon-\epsilon_k} \ = \ v_{i\ell}
+ W'_{i\ell}  \eqno(3.4)
$$
is an energy dependent, but deterministic quantity. Expression 3.1 can be rewritten as an
operator equation.
$$
G \ = \ G^o + G^o W G  \eqno(3.5)
$$
which is valid only for  the two-dimensional adsorbate lattice. Equation 3.5 implies
$$
G \ = \ \frac{1}{\epsilon-\sum_{i\sigma} \epsilon_{i\sigma} n_{i\sigma} -W} \eqno(3.6)
$$
As noted earlier, CPA essentially replaces $\epsilon_{i\sigma}$ in (3.6) by
$k_{\sigma}(\epsilon)$, i.e.,
$$
\sum_{i,\sigma} \epsilon_{i \sigma} n_{i\sigma} \longrightarrow  \sum_{i\sigma}
k_{\sigma}(\epsilon) n_{i\sigma} \equiv K(\epsilon) \eqno(3.7)
$$
Consequently, the configuration averaged Greenian operator is
$$
<G>_c \ \equiv \ \bar{G} \ = \ \frac{1}{\epsilon-K(\epsilon)-W}  \eqno(3.8)
$$
The coherent potential $k_{\sigma}(\epsilon)$ is to be determined self-consistently through
the expression [20]:
$$
\bar{G}^{\sigma}_{ii} \ = \ \frac{1-\theta}{\epsilon_{a\sigma} - k_{\sigma}(\epsilon)}  \eqno(3.9)
$$
In order to determine $k_{\sigma}(\epsilon)$, an explicit expression for $\bar{G}^{\sigma}_{ii}$ is needed.
As the effective medium replacing the random adsorbate layer is periodic, the
configuration-averaged GF matrix elements are diagonal in two-dimensional Bloch
representation. Hence for every site
$$
\bar{G}^{\sigma}_{ii}~ \ = \ ~ <i| \frac{1}{\epsilon-K-W} |i> ~  =  ~ \frac{1}{N_{||}} \sum_u \frac{1}
{\epsilon-k_{\sigma}(\epsilon)-W(\epsilon,u)}  \eqno(3.10)
$$
The summation over momentum $u$ is restricted to the first Brillouin zone of 2D
lattice having $N_{||}$ number of sites. $W(\epsilon,u)$ is the Fourier transform of
$W_{ij}$.
$$ \begin{array}{lcl}
W(\epsilon,u) & = & \sum_j e^{i{\bf u}. {\bf R}_{ji}} [v_{ij} + W'_{ij} (\epsilon)] \\
\\
& = & \epsilon_u + \sum_j e^{i{\bf u}. {\bf R}_{ji}} W'_{ij} (\epsilon) \\
\\
& = & \epsilon_u + W'(\epsilon,u) \end{array} \eqno(3.11)
$$
Using (3.9) and (3.10), the self-consistency equation  for $k_{\sigma}(\epsilon)$ can be written as
$$
\bar{G}^{\sigma}_{ii} \ = \ \frac{1}{N_{||}} \sum_u \frac{1}{\epsilon-k_{\sigma}(\epsilon) - \epsilon_u - 
W'(\epsilon,u)} \ = \ \frac{1-\theta}{\epsilon_{a\sigma} - k_{\sigma}(\epsilon)}  \eqno(3.12)
$$
In the next section, we consider the evaluation of the adsorbate density of states
and its occupation probability using the adsorbate GF $\bar{G}^{\sigma}_{ii}$.

\section {Adsorbate density of states and occupation probability}

~~~~To evaluate the adsorbate density of states at site $i$, we introduce a restricted
configuration averaged GF $<G^{\sigma}_{ii}>_{i=a}$. Here the restricted configuration
average means that we a priori consider the site $i$ to be occupied by an adsorbate.
For the remaining sites, whose occupancy status is left unspecified, 
configuration average is performed. The adsorbate density of states at the site $i$ is
given as:
$$
\rho^{\sigma}_{i} {(\epsilon)} \ = \ \frac{1}{\pi} \rm {Im} <G^{\sigma}_{ii}(\epsilon)>_{i=a}  \eqno(4.1)
$$
The imaginary part of the Green's function is evaluated by taking $\epsilon \equiv 
\epsilon - i0^+$ .
A further configuration average at site $i$ leads to
$$
<<G^{\sigma}_{ii}>_{i=a}> \ = \theta \ <G^{\sigma}_{ii}>_{i=a} \ = \ \bar{G}^{\sigma}_{ii} \eqno(4.2)
$$
From (4.1) and (4.2)
$$
\rho^{\sigma}_{i}(\epsilon)  \ = \ \frac{1}{\pi \theta} \rm {Im}~ \bar{G}^{\sigma}_{ii} \eqno(4.3)
$$
Given $\rho^{\sigma}_{i} {(\epsilon)}$, the average occupation probability is
determined through the relation
$$
<n_{i\sigma}> \ = \ \int^{\epsilon_f}_{-\infty} \rho^{\sigma}_{i} (\epsilon) d\epsilon
~ \equiv ~ < n_{a \sigma} >         \eqno(4.4)
$$
where $\epsilon_f$ is the Fermi energy. As $\bar{G}^{\sigma}_{ii}$ implicitly~ depends on $<n_{i\sigma}>$, a self-consistent
determination of $~<n_{i\sigma}>~$ is required. We determine $~$ it in the non-magnetic limit, 
i.e.,  $~ <n_{i\sigma}>$ = $<n_{i\bar{\sigma}}>$ . Thus, in the present formalism, we need
two self-consistency calculations, one  for the coherent potential ~$k_{\sigma}(\epsilon)$,
and the other for $<n_{i\sigma}>$.

\section{Interaction Energy}
~~~~The interaction energy for the chemisorbed system is defined as
$$
\Delta \bar{E} \ = \ <H> - <H^o>  \eqno(5.1)
$$
where $H$ is the Hamiltonian (2.5), with $\epsilon_{i\sigma}$ replaced by $k_{\sigma}(\epsilon)$.
$H^o$ is the Hamiltonian of the far-separated substrate-adsorbate system.
$$ \begin{array}{lcl}
H^o & = & \sum_{k,\sigma} \epsilon_k n_{k\sigma} + \theta N_{||} \left[ \sum_\sigma E_a
n_{a\sigma} + U n_{a\sigma} n_{a \bar{\sigma}}\right] \\
\\
& + & \sum^4_{\nu =1} \omega_\nu b^{\dagger}_\nu b_\nu + \theta N_{||} \left[ \sum^3_{\nu =1}
(\lambda^o_{a\nu} n_{a\sigma} + \lambda^o_{c\nu})\right] (b_\nu + b^{\dagger}_\nu)
\end{array} \eqno(5.2)
$$
$\lambda^o_{\alpha \nu} (\alpha = a,c~ ;~ \nu = 1,2,3)$ is the coupling coefficient
between the adsorbate in the bulk electrolyte and solvent polarization modes. An
adsorbate far removed from the electrode surface does not experience any image
effect, so we take $\lambda^o_{\alpha 4} =0$. $\lambda^o_{\alpha \nu}$ differs
from the respective $\lambda_{\alpha \nu}$ because  the
solvation energies of an adsorbate, when it is located at the interface and when it is
in the bulk, are different. $\theta N_{||}$ is the total number of adsorbates that 
have migrated from the bulk region to form the 2D adlayer at the interface.

As in section 2, we  again decouple the fermionic and bosonic components in
$H^o$. Subsequent averaging of the bosonic operators [16] in $H$ and $H^0$
leads to
$$
\Delta \bar{E} \ = \ \Delta \bar{E}_1 + \Delta \bar{E}_2 + \Delta \bar{E}_3 
\eqno(5.3)
$$
where
$$
\Delta \bar{E}_1 \ = \ < [ \sum_{k,\sigma} \epsilon_k n_{k\sigma} + \sum_{i,\sigma}
k_{\sigma}(\epsilon) n_{i\sigma} + \sum_{i,k,\sigma} (v_{ik} c^{\dagger}_{i\sigma} c_{k\sigma} +
h.c.) + \sum_{i\ne j, \sigma} v_{ij} c^{\dagger}_{i\sigma} c_{j\sigma} ] > \eqno(5.4)
$$
$$
$$
$$
\\
\Delta \bar{E}_2 \ = \ -< [ \sum_{k, \sigma} \epsilon_k n_{k\sigma} + \theta N_{||} \{
\sum_\sigma \epsilon^o_{a\sigma} n_{a\sigma} +U n_{a\sigma} n_{a\bar{\sigma}}\}] >^o
\eqno(5.5)
$$
$$
$$
$$ \begin{array}{lcl}
\Delta \bar{E}_3 & = & \theta N_{||} \left[ \sum_{\nu =1,2}
\frac{\lambda_{a\nu}^2}{\omega_\nu} (\sum_\sigma <n_{a\sigma}>)^2 - \sum^4_{\nu =1}
\frac{\lambda^2_{c\nu}}{\omega_\nu} \right. \\
\\
& - & (U -  2 \sum^4_{\nu =3} \frac{\lambda_{a\nu}^2}{\omega_\nu}) <n_{a\sigma}>
<n_{a \bar{\sigma}}> \\
\\
& - & \left. \left\{ \sum^2_{\nu=1} \frac{\lambda^{o2}_{a\nu}}{\omega_\nu} (\sum_{\sigma} <n_{a\sigma}>^o)^2 -
\sum^3_{\nu =1} \frac{\lambda^{o2}_{c\nu}}{\omega_\nu} \right\} \right] \end{array}
\eqno(5.6)
\\
$$
$$
$$


$$
\\
\begin{array}{lcl}
\epsilon^o_{a\sigma} \ & = & \ E_a + \sum^2_{\nu =1} \frac{2 \lambda^{o2}_{a\nu}}{ \omega_\nu}
\left(\frac{\lambda_{c\nu}^o}{\omega_\nu} - \sum_\sigma < n_{a\sigma}>^o\right) \\ 
\\
& - & 2 \frac{\lambda^{o2}_{a3}}{\omega_3} < n_{a\bar{\sigma}}>^o 
 +  2 ( \frac{ \lambda^o_{c3} \lambda^o_{a3}}{\omega_3} - \frac{\lambda^{o2}_{a3}}{\omega_3}) \\ 
 \end{array} \eqno(5.7)
$$
$$
$$
$<...>$ and $<...>^o$, respectively, imply the averaging with respect to the states of
interacting and far-removed adsorbate-substrate systems.

Using the Heisenberg equation of motion for Green's functions along with the relation (4.3) and
$$
\rho^{\sigma}_k(\epsilon) \ = \ \frac{1}{\pi} \rm {Im}~ \bar{G}^{\sigma}_{kk}(\epsilon)\,, \eqno(5.8)
$$
$\Delta \bar{E}_1$ and $\Delta \bar{E}_2$ can be expressed in terms of substrate and the
adsorbate layer density of states
$$ \begin{array}{lcl}
\Delta \bar{E}_1 + \Delta \bar{E}_2 & = & \sum_\sigma \left\{
\int^{\epsilon_f}_{-\infty} \epsilon  \rho^\sigma_m(\epsilon) d\epsilon - 
\int^{\epsilon_f^o}_{-\infty} \epsilon \sum_k \rho^{o \sigma}_k(\epsilon) d\epsilon \right\}
\\
\\
& - & \theta N_{||} \left\{ \sum_\sigma \epsilon^o_{a\sigma} <n_{a\sigma} >^o + U
<n_{a\sigma}>^o <n_{a\bar{\sigma}}>^o\right\} \end{array} \eqno(5.9)
$$
where we have used
$$
<n_{a\sigma} n_{a\bar{\sigma}}>^o \ = \ <n_{a\sigma}>^o <n_{a\bar{\sigma}}>^o
$$
for an isolated adsorbate. In this case, $<n_{a\sigma}>^o$ and $<n_{a\bar{\sigma}}>^o$ can
take values 0 or 1 only. $\epsilon_f$ and $\epsilon_f^o$ are the Fermi energies in
the presence and in the absence of chemisorption.
$$
\rho^\sigma_m (\epsilon) \ = \ \sum_k \rho^\sigma_k (\epsilon) + \sum_i
\rho^\sigma_i(\epsilon) \eqno(5.10)
$$
is the total electronic density of states for chemisorbed system.

Using the charge conservation criterion 
$$
\sum_{\sigma} \int^{\epsilon_f}_{-\infty} \rho^\sigma_m (\epsilon) d\epsilon \ = \
\sum_{k, \sigma }\int^{\epsilon^0_f}_{-\infty} \rho^{o \sigma}_{k} (\epsilon) d\epsilon +
\theta N_{||} \sum_{\sigma} < n_{a\sigma}>^o \eqno(5.11)
$$
we have
$$ \begin{array}{lcl}
\Delta \bar{E}_1 + \Delta \bar{E}_2 & = & \sum_\sigma \int^{\epsilon^o_f}_{-\infty}
(\epsilon - \epsilon_f^o) (\rho^\sigma_{m}(\epsilon)- \sum_k \rho^{o\sigma}_k (\epsilon)) d\epsilon \\
\\
& - & \theta N_{||} \{\sum_\sigma (\epsilon^o_{a\sigma} + \epsilon_f^o) <n_{a\sigma} >^o +
U<n_{a\sigma}>^o <n_{a\bar{\sigma}}>^o \} \end{array} \eqno(5.12)
$$
To obtain $\rho^{\sigma}_m(\epsilon)$, we need both adsorbate   (cf. eq 3.12) and the
substrate Green's functions. The latter is obtained from the relation
$$ \begin{array}{lcl}
\bar {G}^{\sigma}_{kk'} & = & \bar{G}^{o\sigma}_{kk} \delta_{kk'} + \sum_{i,j} \bar {G}^{o\sigma}_{kk} v_{ki} \bar {G}^{\sigma}_{ij}
v_{jk'} \bar {G}^{o\sigma}_{kk'} \\
\\
& = & \bar {G}^{o\sigma}_{kk} \delta_{kk'} + \frac{1}{N_{||}} \sum_{i,j} \frac{\bar {G}^{o\sigma}_{kk} e^{i
{\bf u}. {\bf R}_{ij}} v_{ki} v_{jk'} \bar {G}^{o\sigma}_{kk'}}{\epsilon-k_{\sigma}(\epsilon) - W(u,\epsilon)} 
\end{array} \eqno(5.13)
$$
with
$$
\bar {G}^{o\sigma}_{kk} (\epsilon) \ = \ (\epsilon - \epsilon_k)^{-1}
$$
Equations 5.12 and 5.13 allow us to rewrite the interaction energy expression as
$$ \begin{array}{lcl}
\Delta \bar{E} & = & \frac{\displaystyle 1}{\displaystyle \pi} \rm {Im} \sum_\sigma \int^{\epsilon_f}_{-\infty}
(\epsilon - \epsilon_f) (\sum_u \frac{\displaystyle 1-\frac{\displaystyle \partial W(u,\epsilon)}{\displaystyle \partial \epsilon}}{\displaystyle \epsilon-k_{\sigma}(\epsilon) - W(u,\epsilon)})
d\epsilon \\
\\
& - & \theta N_{||} \left \{ \sum_\sigma \epsilon^o_{a\sigma} <n_{a\sigma}>^o + U <n_{a\sigma}
>^o <n_{a\bar{\sigma}}>^o\right \} +
\Delta \bar{E}_3 \,.  \end{array} \eqno(5.14)
$$
The expressions (4.3), (4.4), and (5.14) along with (5.6) constitute  our main results for
the adsorbate density of states, its average electronic charge, and the interaction  energy.

The binding energy of a single adsorbate can be obtained by dividing  $\Delta \bar{E}$
with $\theta N_{||}$.
$$
\Delta E \ = \ \frac{1}{\theta N_{||}} \Delta \bar{E} \eqno(5.15)
$$

Appropriate expressions  suitable for the numerical calculations are
derived in the next section.

\section{Electrochemisorption : A simple model for numerical calculations}

~~~~The results given in the last section require tedious summations over the momentum
 $k$ of the metal states and the momentum  $u$ of the Bloch states in
the 2D adsorbate layer. But we can achieve  considerable simplification under the
following conditions [20].

1.  It is mentioned in the section 2 that the adsorbate layer is commensurate with
the underlying substrate. As a result, the Brillouin zones of the adsorbate layer
and the underlying parallel two-dimensional substrate lattice 
are the same. Assuming the separability of the metal states energies
$\{\epsilon_k\}$ in the directions parallel and perpendicular to the surface, one can
write
$$
\epsilon_k \ = \ \epsilon_{u,k_z} \ = \ \epsilon^{||}_u + \epsilon^{ \perp}_{k_z} 
\eqno(6.1)
$$
where the zero of the energy scale is taken at the substrate band center.

2.  The  density of states for the substrate in the direction
perpendicular to surface is taken to be Lorentzian, whereas the same  
is assumed to be rectangular along the surface :
$$
\frac{1}{N_\perp} \sum_{K_z} \delta (\epsilon -  \epsilon^{\perp}_{k_z})
 \ = \ \frac{1}{\pi} \frac{\Delta_{\perp}}{(\epsilon - \epsilon^{\perp}_{k_z})^2 + \Delta_{\perp}^2}
\eqno(6.2)
$$
$$
\frac{1}{N_{||}} \sum_{u} \delta (\epsilon -  
\epsilon^{||}_u )\ = \ 
\left[ \begin{array}{cc}\frac{1}{2\Delta_{||}} & ~~ -\epsilon < \Delta_{||} <  
\epsilon \\ &  \\ 0 & {\rm otherwise.} \end{array} \right.  \eqno(6.3)
$$
$2\Delta_{||}$ is  the substrate bandwidth  at the surface and
$\Delta_t$, the total bandwidth of the substrate, is given as  
$$
2(\Delta_{||} + \Delta_{\perp}) \ = \Delta_t  \eqno(6.4)
$$
$N_\perp$ is the number of atomic layers in the substrate in the direction
perpendicular to its surface.

3. An adsorbate occupying the ``on-top'' position on the substrate is assumed to
couple with the underlying substrate atom only. Therefore,
$$
v_{ik} \ = \ \frac{1}{\sqrt {N_{\perp} N_{||}}} v e^{i{\bf k} . {\bf R}_i} 
\eqno(6.5)
$$
where $v$ is the ``adsorbate - nearest substrate atom'' coupling strength.

4.  As the adsorbate layer is commensurate with the underlying substrate layer,
both these layers have identical geometrical configurations.
The matrix element $v_{ij}$
depends on the distance $R_{ij}$ and a universal constant specific to the system [24].
Since $R_{ij}$ is same for the adsorbate and substrate layers, 
the hopping term $v_{ij}$ for the adsorbate layer is proportional to
$v_{ij}'$, the hopping term in the substrate's surface layer. 
Taking  $\mu$ as the proportionality constant, we have
$$
v_{ij} \ = \ \mu v'_{ij} \eqno(6.6)
$$
and hence, using the Fourier transform (cf. eq 3.11)
$$
\epsilon_u \ = \ \mu \epsilon^{||}_u  \eqno(6.7)
$$
Since the half-bandwidths of adsorbate layer, $\delta$, and substrate surface,
$\Delta_{||}$, are proportional to $v_{ij}$ and $v'_{ij}$ respectively, we can
write
$$
\mu \ = \ \frac{\delta}{\Delta_{||}}  \eqno(6.8a)
$$
The half-bandwidth $\delta$ in the  above relation corresponds to the case of
a complete monolayer. But, in fact, the bandwidth of adlayer depends on the coverage
and tends to zero in the low-coverage limit. Following the methodology described
in ref 22, we can write $\delta (\theta)$, the coverage-dependent half-bandwidth 
of the adlayer, as $\theta \delta$. Consequently, for a general coverage, eq 6.8a
gets modified as
$$
\mu \ = \ \frac{\theta \delta}{\Delta_{||}}  \eqno(6.8b)
$$
It may be noted here that when the coverage tends to zero, the configuration averaged 
adsorbate Green's function (cf. eq 3.10 equals $\theta$ times  the adsorbate GF 
obtained for the ``lone-adsorbate'' case. This remains valid for  any strength   
of the $v_{ij}$ [20,23]. 

~~~~For a given material, if the  total bandwidth $\Delta_t$ is known, the bandwidth
$2\Delta_{||}$ for its surface can be obtained using the relation [25]
$$
2\Delta_{||} \ = \ \frac{N_s}{N_b} \Delta_t \eqno(6.9)
$$
$N_s$ and $N_b$ are the respective co-ordination numbers for the surface and bulk
atoms. Using eqs 3.12 and 6.1-6.7, the adsorbate GF and the self-consistency
equation for the coherent potential $k_{\sigma}(\epsilon)$ can be rewritten as
$$
 \begin{array}{lcl}
\bar{G}_{ii}^\sigma (\epsilon) & = & \frac{ \displaystyle 1-\theta}{ \displaystyle \epsilon_{a\sigma} - k_\sigma (\epsilon)} \\ 
\\
& =  & \frac{ \displaystyle 1}{\displaystyle 2\Delta_{||}  (B - A) \mu} \left[ (A-C) \ln \left(
\frac{A-\Delta_{||}}{A+\Delta_{||}}\right) - (B-C) \ln \left(
\frac{B-\Delta_{||}}{B + \Delta_{||}}\right) \right] 
\end{array} \eqno(6.10)
$$
with 
$$
A/B \ = \ \frac{1}{2} [(C+D) +/- \{(C+D)^2 - 4(CD - \frac {v^2}{\mu}) \}^{\frac{1}{2}}] 
\eqno(6.11)
$$
$$
C \ = \ \frac{\epsilon - k_\sigma(\epsilon)}{\mu} \quad; \quad \ D \ = \ \epsilon - i \Delta_{\perp}
\eqno(6.12)
$$
Similarly, the interaction energy (cf. eqs 5.14 and 5.15) can be expressed as 
$$ \begin{array}{lcl}
\Delta \bar {E} & = & \frac{\displaystyle N_{||}}{\displaystyle \pi} \rm {Im} \sum_\sigma \left \{ \int^{\epsilon_f}_{-\infty}
(\epsilon - \epsilon_f) \{\bar{G}^\sigma_{ii} + \xi^\sigma \} d\epsilon \right \}
 \\
\\
& - & \theta N_{||} \{\sum_\sigma (\epsilon^o_{a\sigma} + \epsilon_f) <n_{a\sigma} >^o + U
<n_{a\sigma}>^o <n_{a\bar {\sigma}
}>^o \} 
 +  \Delta \bar{E}_3 \end{array} \eqno(6.13)
$$
$\Delta \bar{E}_3$ and $\epsilon^o_{a\sigma}$ are given in eqs 5.6 and 5.7, and 
$$ \begin{array}{lcl}
\xi^\sigma & = & \frac{1}{2\Delta_{||}} \left[ \ln \left(\frac{D- \Delta_{||}}{D +
\Delta_{||}}\right) + \frac{1}{B-A} \left\{(A-C) \ln \left(
\frac{A-\Delta_{||}}{A+\Delta_{||}}\right) - (B-C) \ln
\left(\frac{B-\Delta_{||}}{B+\Delta_{||}}\right) \right\}\right]
\end{array} \eqno(6.14)
$$

\noindent The expression 6.10 for the adsorbate GF contains  $\mu$ in the denominator.
Though $\mu$ tends to zero when the $\theta$ approaches zero,
the $\bar{G}^{\sigma}_{ii}(\epsilon)$
remains finite in this limit. This can be verified either by suitable expansion of
adsorbate GF, or by a priori taking $v_{ij}$ (and hence $\mu$) to be zero and
reevaluating the adsorbate Green's function [20,23].

In the above derivations  for the adsorbate GF and the binding energy, $W'(\epsilon, u)$,
the Fourier coefficient of $ W'_{ij}$ (cf. eqs 3.4 and 3.11), has been explicitly evaluated 
using the model density of states (eqs 6.2 and 6.3) and the relation 6.5. Alternatively,
$\bar{G}^{\sigma}_{ii}$ and $\Delta \bar{E} $ can be evaluated using the wide-band approximation.
As noted in the Introduction, this approximation has been extensively used in the
chemisorption problems  at electrochemical interface, as well as in the Kravstov and
Mal'shukov work on adlayer [21]. Accordingly, we now have 
$$
W'_{il}(\epsilon) ~ \approx ~ i\Delta    \eqno(6.15)
$$
where $\Delta$ is an energy-independent quantity and is a measure of adsorbate's energy
width [17,21]. Within the above approximation, expression 6.10 for the adsorbate
GF now becomes 

$$
 \begin{array}{lcl}
\bar{G}^{(w) \sigma }_{ii} (\epsilon) & = & \frac{ \displaystyle 1-\theta}{ \displaystyle \epsilon_{a\sigma} - k_\sigma (\epsilon)} \\ 
\\
& =  & \frac{ \displaystyle 1}{\displaystyle 2\Delta_{||}   \mu}  \ln \left(
\frac{ \displaystyle \epsilon -k_{\sigma}(\epsilon) + \mu \Delta_{||} - i\Delta}{\displaystyle \epsilon -
k_{\sigma}(\epsilon) - \mu  \Delta_{||} - i \Delta } \right) 
\end{array} \eqno(6.16)
$$

In the $ \theta \rightarrow 0 $ limit, the $ \bar{G}^{(w) \sigma}_{ii}$ reduces 
to $\theta$ times the adsorbate GF $ G_{aa}(\epsilon)$  obtained by Kornyshev and Schmickler
[17,20].  

$$
\bar{G}^{(w) \sigma}_{aa}(\epsilon)  =  \frac {\displaystyle \theta}
{ \displaystyle \epsilon - \epsilon_{a\sigma} - i\Delta }  \eqno (6.16a)
$$

The superscript $(w)$ here specifies the wide-band limit.
While calculating the average adsorbate charge and the binding energy,
the $\theta$ appearing in numerator of the right-hand side of the above expression gets
canceled with a $\theta $ in the denominator (cf. eqs 4.3 and 5.15) and we recover the
lone-adsorbate result of  Kornyshev and Schmickler.

When the interaction energy is evaluated within the
wide-band approximation, the
$ \xi^{\sigma}$ appearing on the right-hand side of eq 6.13 is identically zero. 
Also the integral in this expression can be rewritten as (cf. eqs 4.3 and 4.4):

$$ \begin{array}{lcl}
\frac{\displaystyle 1}{\displaystyle \pi}\rm {Im}\int^{\epsilon_f}_{-\infty}
(\epsilon - \epsilon_f) \bar{G}^{(w)\sigma}_{ii} d\epsilon & = & 
 \frac{\displaystyle 1}{\displaystyle \pi}\rm {Im}\int^{\epsilon_f}_{-\infty}
(\epsilon_{a\sigma} +(\epsilon - \epsilon_f - \epsilon_{a \sigma})) \bar{G}^{(w)\sigma}_{ii}
d \epsilon \\
\\
& = & \theta ~ \epsilon_{a\sigma} < n_{i \sigma} > + \frac{\displaystyle 1}{\displaystyle \pi}
\rm {Im}\int^{\epsilon_f}_{-\infty}( \epsilon -
\epsilon_f - \epsilon_{a\sigma}) \bar{G}^{(w) \sigma}_{ii} d \epsilon \\ 
\end{array} \eqno(6.17)
$$
The last integral in the above expression diverges, and its lower limit needs a cutoff 
for obtaining a finite result. This energy cutoff is put at  $\epsilon _ {\Gamma}$, 
the bottom of conduction band in the metal [18]. Thus, in the wide-band limit, the
interaction energy expression (6.13) gets replaced by 

$$ \begin{array}{lcl}
\Delta \bar {E}^{(w)} & = &  N_{||}  \sum_\sigma \left \{ \theta \epsilon_{a \sigma} <
n_{a \sigma}>  + \frac{\displaystyle 1}{\displaystyle \pi} \rm {Im} \int^{\epsilon_f}_{\epsilon_{ \Gamma}}
(\epsilon - \epsilon_f - \epsilon_{a \sigma}) \{\bar{G}^{(w)\sigma}_{ii}   \} d\epsilon \right \}
 \\
\\
& - & \theta N_{||} \{\sum_\sigma (\epsilon^o_{a\sigma} + \epsilon_f) <n_{a\sigma} >^o + U
<n_{a\sigma}>^o <n_{a\bar {\sigma}
}>^o \} 
 +  \Delta \bar {E}_3 \end{array} \eqno(6.18)
$$

In  the following  sections, the details of the electronic structure of an 
copper layer adsorbed on gold electrode are considered. We give  the results both for 
the general analysis and the wide-band approximation. A comparative study of these two 
approaches for the chemisorption problem is also provided. 

\section{Electrochemisorption of copper on gold electrode}

~~~~The relevance of substrate d band in the chemisorption phenomenon is well recognized.
The importance of d band arises from its high density of states. Besides, the
substrate d-orbitals provide strong, directed bonding with the adsorbate electronic
states. Consequently, we model the gold substrate through its d band in the present
calculations. But it may be mentioned here that the gold d band lies deeply
under the Fermi level. In fact the gold Fermi level lies inside the sp-band. Therefore, one
needs to include both d and sp bands of gold  in the chemisorption calculations.
In this regard, the Lorentzian density of states in eq 6.2 can be considered
as the effective density of states which includes both d and sp bands of gold.
It may be noted that the density of  states for d band alone is semielliptical,
and not Lorentzian.

In an earlier calculation for the electrosorption, the participation
of only the s orbital of copper in the bonding was assumed [26]. But a proper analysis requires the
inclusion of both s and d orbitals of copper in the model.
There are three main reasons why a d-orbital of copper is also relevant to electrosorption
process: (i) energetically it lies near the substrate d band when the intraadsorbate
Coulomb repulsion $U_d$ is taken into account, (ii) it has a large matrix element with the
substrate  states, and (iii) during the chemisorption of $Cu^{2+}$, a substantial amount of
charge transfer occurs from the substrate to an unoccupied copper d-orbital. Therefore,
both s and d orbitals of copper are included in the  calculations. Subsequent
analysis then requires a simultaneous self-consistent determination of the adsorbate
charges $<n_{s\sigma}>, <n_{d\sigma}>$, and the coherent potentials
$k_{s \sigma }(\epsilon)$ and $k_{d \sigma}(\epsilon)$. In order to simplify this
difficult task, we neglect the indirect interactions between s and d orbitals. Thus
we treat the electrosorption of s and d orbitals separately. The self-consistency
criterion is now applicable to individual pairs $(<n_{s\sigma}>,
k_{s \sigma}(\epsilon))$ and $(<n_{d\sigma}>, k_{d \sigma}(\epsilon))$ only.

\subsection{System Parameters}

~~~~The energies of copper s and d orbitals lie at 7.72 eV and 20.14 eV below the vacuum
level [24]. In the present calculations, energy zero is taken at the center of the
substrate d band. Hence
$$
E_s \ = \ \phi + \epsilon_f - 7.72 \quad \ ; \ \quad E_d \ = \ \phi + \epsilon_f - 20.14
\eqno(7.1)
$$
$\phi$ is the substrate work function  and $\epsilon_f$  is  its Fermi level
measured from the band center. The intraadsorbate repulsion energy for s and d
orbitals, namely $U_s$ and $U_d$, are 5.92 eV and 5.45 eV, respectively [24,27].

We employ the Dogonadze, Kornyshev, and Schmickler scheme [26,28] to evaluate the
contributions of various solvent polarization modes toward the total solvation
energy. Accordingly, the contributions from the orientational, vibrational, and
electronic polarization modes are respectively given as:
$$ \begin{array}{lcl}
E_1 & = & \alpha X (\frac{\displaystyle 1}{\displaystyle \epsilon_{ir}} - \frac{\displaystyle 1}{\displaystyle \epsilon_s})
F(\lambda_1/r_i) \\
\\
E_2 & = & \alpha X (\frac{\displaystyle 1}{\displaystyle \epsilon_{opt}} - \frac{\displaystyle 1}{\displaystyle \epsilon_{ir}})
F(\lambda_2/r_i)  \\
\\
E_3 & = & \alpha X (1- \frac{\displaystyle 1}{\displaystyle \epsilon_{opt}})
F(\lambda_3/r_i) \end{array} \eqno(7.2) 
$$
where 
$$
F(x) \ = \ 1-(1-e^{-x})/x \eqno(7.3)
$$
and the correlation lengths $\lambda_1, \lambda_2,$ and $\lambda_3$ for the three
polarization branches are 0.68, 0.12, and 0.10 nm respectively.  The ionic radius is 
$r_i$ (0.127, 0.096, and 0.072 nm for $ Cu^0$ , $Cu^{1+}$ and $Cu^{2+}$, respectively) 
and $\epsilon_{ir}$ (= 4.9), $\epsilon_{opt}$ (1.15), and $\epsilon_s$ (78.5)
are the infrared, optical, and static dielectric constants for water.
The radius of a fractionally charged adsorbate is obtained through the interpolation
formula provided in ref 26.
$\alpha$ is the degree of solvation of the adsorbate at the interface.
The parameter $X$ is so chosen that $ z^2 (E_1 + E_2 + E_3)/ \alpha$
equals the solvation energy $\Delta G_S$ of an ion having charge $z$. $X$
in fact accounts for any discrepancy in the calculated and experimental
solvation energies. At a metal-vacuum interface,  the image energy of a unit charge is taken as $-e^2/4
(x+ \kappa^{-1})$ [29], where $\kappa^{-1}$ is the Thomas-Fermi screening length.
For gold, $\kappa^{-1}$ = 0.06 nm. $x$ is the distance of the charge from the
electrode surface. We take the  distance
between the copper adsorbate and the underlying gold atom to be the same as
the distance between two nearest adsorbed copper atoms in the monolayer
configuration, i.e., 0.288 nm [4]. The electrode plane is considered to
be located at the midpoint between the copper and gold atoms. Therefore,
the distance $x$ between the adsorbate and substrate surface is 0.144 nm.
Substituting these values, we get the image energy $\epsilon_{im} = -1.76~ eV$. 
Thus, we now have
$$
\frac{\lambda^2_{a \nu}}{\omega_{\nu}} \ = \ E_{\nu} ~~ (\nu=1,2,3 \quad {\rm unless \
specified \ otherwise)} \eqno(7.4)
$$
and at the M-V interface
$$
\frac{\lambda^2_{a4}}{\omega_4} \ = \ - \epsilon_{im} \ = \ 1.76~ eV\     \eqno(7.4a)
$$

At an electrochemical interface, we use the Kornyshev and Schmickler formalism 
for determining the image energy [26].

$$
\ - \frac{\lambda^2_{a4}}{\omega_4} \ = \epsilon_{im} \ = \  \frac{\displaystyle e^2_o}
{\displaystyle 4 \epsilon_{opt} r_i} \frac {\displaystyle (\epsilon_{opt} - 1) - \kappa^2
r^2_i} {\displaystyle (\epsilon_{opt} + 1 ) + \kappa^2 r^2_i}  \eqno(7.4b)
\\
$$

As noted earlier, $(E_1 + E_2 + E_3)/ \alpha  \ = \ \Delta G_s =
5.92~ eV$ is the solvation energy of a singly charged copper ion [26].
When the adsorbate is in the bulk solution region,
$\lambda^{o2}_{a \nu}/ \omega_{\nu} = E_{\nu}/\alpha$.

The solvation energy of the interfacial adsorbate, as considered above,
is appropriate for  low-coverage regime. At the higher
coverages, adsorbates progressively get desolvated. This is caused by  
(i) steric effects and (ii) delocalization of adsorbate electron over a
large spatial region for metallic adsorbates. The  delocalization of electrons
causes a  weakening of the adsorbate-solvent interaction 
and it also leads to an effective  screening of the adsorbate core
charge. In fact when $\theta = 1$, the electrons are delocalized over the 
entire 2D copper layer. This layer can be now considered to form the
outermost surface of the electrode. Consequently, its interaction with
the solvent polarization, as a first  approximation, can be
neglected. Since no theoretical study of the dependence of the adsorbate  
solvation energy $\Delta G^a_s$  on coverage seems to be available, we propose the following
interpolative scheme  for $E_{\nu}$, and hence for $\Delta G^a_s(\theta)$, as a
function of coverage
$$
E_{\nu}(\theta) \ = \ (1-\theta^2) E_{\nu} \eqno(7.5)
$$
The presence of $\theta^2$ in a sense takes into account the correlation
in a primitive way. 
The decrease  in adsorbate solvation energy 
$E_{\nu}(\theta)$ depends on the probability of finding another
adsorbate in its vicinity.
The relation 7.5 satisfies  the limiting cases
corresponding to $\theta \rightarrow  0$ and $\theta = 1$.  We now have  
$$
\frac{\lambda^2_{a \nu}(\theta)}{\omega_\nu} \ = \ E_{\nu}(\theta) \ = \ (1-\theta^2) E_{\nu} \ = \
(1-\theta^2) \frac{\lambda^2_{a \nu}}{\omega_{\nu}} \eqno(7.6)
$$
$$
\sum_{\nu=1}^3 E_{\nu}(\theta) \ = \  \Delta G^a_s(\theta) \ = \
\alpha(1-\theta^2) \Delta G_s \eqno(7.7)
$$
In the calculation, we take $\lambda_{a{\nu}s} = \lambda_{a{\nu}d} =
\lambda_{a\nu}$. The suffixes s and d denote s and d orbitals. We take the
coupling strengths of a unit positive core charge and the orbital
electron with solvent modes to be of the equal magnitude. In the neutral
state, the copper s orbital is singly occupied, hence 
for  the underlying core charge,  $|\lambda_{c \nu s}|$ is equal to
$|\lambda_{a \nu}|$., On the other hand,
a d orbital is doubly occupied in the neutral state and so
$|\lambda_{c \nu d}| = 2|\lambda_{a \nu}|$.

The ab initio calculation  of the matrix elements $\{v\}$ ($cf.$ eqs 6.5 and 6.11) between the  
orbitals of copper adsorbate and gold electrode  is a difficult task.
We treat them as adjustable parameters in our calculations. Reasonable estimates for
these parameters can be obtained through the existing data .
The $v_s$ for the hydrogen adsorption on transition metals is $\approx$ 4.00 eV [13,23].
Now $v_s$ varies as $R^{-3}$, where $R$ is the inter-atomic spacing [24]. The equilibrium
distance $R_e$ for  an hydrogen atom chemisorbed on copper, or other transition metals is $\sim0.20 nm $
[25]. This is less than 0.288 nm , the separation between copper and the gold substrate. The
$R^{-3}$ dependence of $v_{s}$ implies that for the present system, $v_{s}$ is
smaller than 4.0 eV. In our calculations, we have taken $v_{s}$ as 2.70 eV at the M-V
interface and allowed for its variation at electrochemical interface.
Schrieffer et al. have specified that for transition metals, $v_{d}$ lies in the range 4
$\sim$ 6 eV [29]. We take $v_{d}$ = 4.85 eV. As subsequent calculations will
demonstrate, these specific values of $v_{s}$ and $v_{d}$ enable us to recover
the observed trends in the average adsorbate charge and give reasonable estimates
of the binding energy both for the metal-vacuum and electrode-electrolyte
interface.

Gold forms a fcc lattice. The coordination number $N_b$ of a  gold atom in the bulk is 12. The
coordination number $N_s$ of an  atom at the Au(111) surface is 6. $\Delta_t$, the total d 
bandwidth of gold, is 5.6 eV [24]. Using eq 6.9, we get $\Delta_{||}$ = 1.40 eV. From (6.4)
$\Delta_{\perp}$ = 1.40 eV. The  d bandwidth of copper is 3.47 eV [24]. Through  similar
argument, we get $\delta$, the half d bandwidth  of 2D adsorbed copper layer
as 0.87 eV. Hence from eq 6.8, $\mu_d$ = 0.6 for the copper monolayer. The bandwidth of unhybridized copper s band
is taken as $E(\Gamma)- E(X)$ which equals 4.21 eV [30]. Consequently $\mu_s$ =
0.75. The work function of gold electrode is 4.8 eV, and its Fermi level, measured
from the d band center, is 5.20 eV [24]. We have taken $\alpha$, the degree of solvation of 
the adsorbate,  as 0.50. This value lies exactly at the midpoint of the prescribed range
0.33-0.66 [26]. Finally, the energy $\epsilon_{\Gamma}$ of the  conduction band bottom
for the gold is taken at 8.10 eV below the Fermi energy [22]. 
We recall that this quantity is needed to  calculate of binding energy in the 
wide-band approximation.

Unless stated otherwise, the
various parameters employed in the calculations are summarized in Table 1.

\section{Numerical Results and Discussions}
\subsection {Average occupation probability and energetics}

~~~~With a view to have a comparative study of chemisorption at the metal-vacuum and
electrochemical interface, we begin our calculations with  the same values of the matrix 
elements $v_s (2.70 eV)$ and $v_d(4.85 eV)$ for both the interfaces.

Two specific features concerning the copper d orbital occupancy $<n_{d \sigma}>$ emerges through
the calculations: (i) this orbital is almost completely filled upon chemisorption and,
(ii) the variations in the coverage have little effect on $<n_{d\sigma}>$. This is true
both for the metal-vacuum ($<n_{d\sigma}>  ~\approx~$ 0.97) and electrochemical ($<n_{d\sigma}> 
~\approx ~$  0.965) interfaces. 
Since the level shift caused by external applied potential is not
considered in the calculations, the results reported in this section pertain to the
potential of zero charge.
The plot for self-consistent evaluation of $< n_{d\sigma}>$ at electrochemical interface
is given in Figure 1. The point of interaction between the straight line and 
the curve describing the calculated values  of charge on copper d orbital 
$ N_{d\sigma}$, corresponding to various  input of $ < \bar{n}_{d\sigma} > $ 
(cf. eq 4.4), gives the required self-consistent value.

In contrast to $<n_{d\sigma}>$, the average occupancy $<n_{s\sigma}>$ for copper
s orbital is more sensitive to the coverage variation.                 
As the  Cu d orbital is almost  completely filled, 
any variation in the adsorbate net charge arises  through 
the changes in the $<n_{s \sigma}>$. We first consider the case of metal-vacuum interface, 
wherein $<n_{s\sigma}>$ is around 0.3. The increase in coverage results in a
slight increase in $<n_{s\sigma}>$ (cf. Figure 2). The change is minimal in the 
low-coverage regime. The curve a in  Figure 3 leads to the  self-consistent  evaluation of
$<n_{s \sigma}>$ when $\theta = 0.1$. The  $U$ term 
in the adsorbate's s orbital energy expression leads to an upward level shift with the  
increasing value of input $< \bar{n}_{s \sigma} >$, which in turn reduces the calculated
value of adsorbate charge $ N_{s \sigma}$.
We get a single self-consistent
solution for $< n_{ s \sigma} >$ in the  entire coverage range.
The variation in the copper s orbital binding energy $\Delta E_s $
with respect to $<\bar{n}_{s\sigma}>$ gives rise to an
energy extremum at the self-consistent value of the orbital charge. 
For $\theta $ = 0.9, and the corresponding self-consistent $<n_{s\sigma}>$ at 0.327,
the occurrence of the  extremum in  $|\Delta E_s|$ is shown in Figure 4. Note that
in the present case, $\Delta E_s$ equals  -$|\Delta E_s|$.

For the  chemisorption at the metal-vacuum 
interface, the $\theta$ variation has a minimal effect on 
d orbital binding energy $\Delta E_d$. The major
contribution to the total binding energy $\Delta E$ (= $\Delta E_s + \Delta E_d$) comes 
through $\Delta E_s$ ($\approx$ -1.6 eV). The 
$\Delta E_d$ is $\approx$~ -0.42 eV and the total binding energy  
is $\approx$~ -2.00 eV.

Before taking up the evaluation of  Cu s orbital occupancy at the gold electrode
in the electrochemical environment, let us restate some relevant experimental facts.
The partial charge transfer coefficient $\lambda$ is one of
the important parameters in the electrosorption studies. It equals to the difference
of charges on the adsorbate when it is in the bulk, and when it lies at the
interface. For the chemisorption of ionic $Cu^{2+}$,
$$
\lambda \ = \ 2(<n_{s\sigma}> + <n_{d\sigma}>) -1        \eqno 8.1
$$
Though $\lambda$ itself is not a measurable quantity, its numerical value lies
close to the experimentally determined electrosorption valency $\gamma_n$ [11,12]. In
the $\theta \rightarrow 0$ limit, $\gamma_n$ for Cu/Au system is 2.0 [12].
Thus the  double layer studies coupled with the thermodynamic analysis
indicate that the  copper adsorbed at a gold electrode  is neutral
when the  coverage  is small. 

On the other hand, Tadjeddine et al. have
shown using the  in situ X-ray absorption spectroscopy that in the higher
coverage domain, charge on the copper adatom on a gold electrode is close to +1.0 [6].

This substantial reported  variation in $< n_{ s \sigma}> $ with changing coverage
prompts us to make a simple calculation for $\Delta E^{\prime}_s$, the energy  
required to bring a single $ Cu^{+}$
ion from the bulk electrolyte to the interface when the bonding is
very weak $( v_s \rightarrow 0)$, but the occupancy of the copper's  s  spin-orbital   
in the interfacial region is $<n_{s \sigma}>$. In this case of negligible
bonding, the main contribution to the energetics would come  from  (i) partial desolvation
of the ion at the electrode, (ii) image energy contribution, (iii) change in energy
while transferring charge 2$<n_{s \sigma}>$ from the Fermi level of the electrode
to the unbroadened orbital  of the adsorbate at the interface, and (iv) intraadsorbate
coulomb repulsion $U_s$.
In the case $<n_{s \sigma}> = <n_{ s \bar {\sigma}}> = 0 $, the charge on Cu at the interface 
remains +1. Therefore the energy $ \Delta E^{\prime}_s$ is 
-$ \sum^4_{\nu =1}\lambda_{a\nu}^2/\omega_\nu + \Delta G_s$. As noted earlier, we have taken
$ \lambda_{ac \nu} = \lambda_{a \nu}$. Substituting the numerical values, we
get $ \Delta E'_s~$ = + 1.29 eV. Next, when  $<n_{s\sigma}> = <n_{s \bar{\sigma}}> = 0.5$,
i.e., the copper adsorbate is neutral ( each spin-orbital being  half occupied),
$\Delta E^{\prime}_s$ = $ \phi - I + U_s/4 
- 0.5 \sum^4_{\nu =3}\lambda_{a\nu}^2/\omega_\nu + \Delta G_s$. 
Taking the first ionization energy $I$ for copper as 7.72 eV, we get $\Delta E^{\prime}_s$ as +3.53 eV
for a neutral copper atom at the interface. Finally, when $<n_{s\sigma}> = <n_{s \bar
{\sigma}}> = 1.0 $, that is, the copper atom  at the electrode has a net charge -1.0,
$\Delta E^{\prime}_s$ = $2(\phi - I) + U_s -
\sum^4_{\nu =1}\lambda_{a\nu}^2/\omega_\nu + \Delta G_s$,  which is equal to +1.41 eV.
Note that the analysis given here  corresponds to the lone-adsorbate case;
the influence of the other adsorbates has been neglected. We have taken the degree of
solvation $\alpha $ of the copper ion at the interface as 0.5.
The expressions given here for $\Delta E^{\prime}_s$  for various cases  can be obtained either
directly, or taking $v_s = 0~ $ appropriately in eqs 5.3-5.7 and 5.15. These elementary 
calculations  show  that in the absence of bonding $(~v_s = 0 )$,  bringing a $Cu^{+}$ ion 
from the  bulk phase to the interface is energetically unfavorable, irrespective of
the charge state of copper at the interface.
This remains true for entire plausible range of $\alpha$ between 0.33 and 0.66.
Next the ionic configurations  of copper at the interface have relatively lower energy
than the neutral copper atom, with $ Cu^{+}$ being the least energy configuration. 

In the above calculations, we have taken $v_s$ to be zero. But in order to have different
charge states of copper at the interface, we need a  finite $v_s$, howsoever 
infinitesimal its magnitude may be. In addition, it has to be verified that we indeed get 
more than one self-consistent solutions for $<n_{s \sigma}>$, so that all the charge states 
of copper considered above are  realized at the interface. Taking $v_s$ = 0.02 eV
(which is 2 orders of magnitude lower than its earlier value
of 2.70 eV), and $\theta = 0.1$ (leading to very low concentration of the 
adsorbates), we  get all the three charge states of copper adsorbate.
The self consistent values of $<n_{s \sigma}>$ are 0.0, 0.504, and 1.0.
Their respective binding energies (cf. eqs 5.15 and 6.13) are 1.32, 3.51, and 1.50 eV, which 
compare well with the $\Delta E^{\prime}_s$ values obtained earlier.

Next when we allow $v_s$ to take the  value  2.70 eV, it is pertinent
to examine whether  (i) one  still gets three different charge states of copper, (ii)
the binding energy becomes negative, and (iii) a less ionic state of copper adsorbate is
more stable? The calculations corresponding to $\theta $ = 0.1 confirm the first
point but the answers to  the next two queries are in the negative. The self-consistent
values of $<n_{s\sigma}>$  are now 0.074, 0.665, and 0.98. The associated 
s orbital binding energies  $\Delta E_s$
are -2.32x$10^{-2}$, 1.85, and 1.34 eV respectively. This shows that even a strong coupling 
between the copper s orbital and the metal states does not  make  neutral copper atom
relatively more stable than the positively charged copper adsorbate in the 
low-coverage regime, nor does it lower the energy sufficiently so as to  make  
the chemisorption of copper at gold electrode feasible. As we shall shortly show,
the adsorption is possible due to the bonding of copper d orbital.
Subsequent calculations with progressive increasing  coverage show that
the roots of $<n_{s\sigma}>$ lying at 0.074 and 0.665 shift toward higher values,
whereas the one at 0.98 moves downward. At $\theta$ = 0.52 the lowest root
shifts to 0.09 ($\Delta E_s$ = 0.52 eV) and  the remaining two roots merge and take 
a common value 0.89 ($ \Delta E_s $ = 2.01 eV). Beyond this coverage, we get only 
one self-consistent solution for $< n_{s\sigma}>$ giving rise to a positively
charged copper adsorbate. Thus when $\theta $ = 0.75, $<n_{s\sigma}>$ = 0.14
( $\Delta E_s $= 1.02 eV) and the same for $\theta$ = 1.0 is 0.33
($\Delta E_s$ = 1.23 eV).

When $\theta$ = 0.1 and $v_s$ = 2.70 eV,
the net charge Q on copper corresponding to various self-consistent 
values of $<n_{s\sigma}>$ is +0.92, -0.26, and -0.89. These values are obtained by
adding the  d orbital and core charges to 2$<n_{s\sigma}>$.

$$
Q  = - 2 ( <n_{s\sigma}> + <n_{d \sigma}> ) + 3            \eqno 8.2
$$

The least charge on copper
is -0.26, which does not make the Cu adatom exactly neutral. In fact it has been shown
earlier that the neutral state of copper is obtained when  $v_s$ =0.02 eV.
Thus, in order to lower the relative large negative charge -0.26, we now take
$v_s$ = 2.0 eV. The charge -0.26 now gets reduced to -0.11. The other motivation
for taking $v_s$ as 2.0 eV, apart from its earlier considered value of 2.70 eV, is
that it   allows us to understand the effect of $v_s$ variation on chemisorption
characteristics.  We also note that the variation of $v_d$ has least effect on the d
orbital occupancy. 

The self-consistent plot  for $<n_{s\sigma}>$ when $v_s$ = 2.0 eV is given in
Figure 3. For $\theta = 0.10$, we get three self-consistent values at  0.04, 0.59,
and 0.99. The change in the locations of  various roots with increasing coverage 
follows a trend identical to the $v_s$ =  2.70 eV case. But instead of at $\theta$
= 0.52, the merger of the two higher roots takes place at $\theta$ = 0.65. 
At this point, the common value of these roots are 0.88. The remaining root lies at
0.079. Beyond $\theta$ = 0.65, only the smallest  root survives, whose magnitude
keeps on increasing with further increase in coverage. Finally, at $\theta$ = 1.00,
the  $<n_{s\sigma}> $ is 0.39 which is 0.06 more than
the corresponding value when $v_s$ = 2.70 eV.

It may be noted that the slow boson terms, weighted by $< n_{s \sigma} >$,
now lower the adsorbate orbital energy with increasing $ < n_{s \sigma} >$  (cf. eq 2.6).
As the magnitude of these quantities are greater than the $ U $ term (unless $ \theta $ is 
quite large (cf. eq 7.7)), calculated $ N_{s \sigma}$ increases with increasing
value of input $< \bar {n}_{s \sigma} >$. 

The  occurrence of extremum in $|\Delta E_s|$ for $\theta$ = 0.1 ($<n_{s\sigma}>$ =
0.04, 0.59 , 0.99) and $\theta$ = 0.67 ($<n_{s \sigma}>$ = 0.082) at electrochemical 
interface is shown in Figure 4. Note that presently  $ \Delta E_s$ = $|\Delta E_s|$.

The net charge on copper with varying coverage
at the electrochemical 
interface is shown in Figure 5. Since the
d orbital occupancy is virtually constant, the structure
in charge variations  solely arise due to $<n_{s\sigma}>$. 
At the gold electrode, 
we get three different charged states of Cu up to $\theta$ = 0.65. The states which are
almost completely ionic when $\theta$ is small, tend to lower their net charge
when coverage increases. This variation is relatively  faster for the positive charge
state. On the other hand, the $Cu^{-0.11}$ configuration in the low-coverage region
progressively acquires more electronic charge with increase in $\theta$. But
beyond a critical coverage ($\theta = 0.65$), both the negatively charged 
configurations  of copper cease to exist. Only the state having positive  charge,
which is also the least energy configuration for all $\theta$ values, survives.
Given the net charge Q on the copper adsorbate, the partial charge transfer coefficient  
can also be obtained using the relation $\lambda~=~2 - Q$ (cf. eqs 8.1 and 8.2).

It is pertinent to analyze the reason behind the transition  from multiple solutions to a 
single root for $<n_{s\sigma}>$ with increasing coverage at electrochemical interface.
We obtain only a single  self-consistent 
solution for  the adsorbate charge at metal-vacuum interface in the entire coverage 
( 0 $<$ $\theta  \le 1 $) range. The multiple roots  occur only at electrochemical interface,
where the solvation of adspecies and  its progressive desolvation with increasing
coverage are the additional features. We note that  presently  
the effect of external applied potential has not been considered. 
It is the presence of  additional boson modes corresponding to the solvent
polarization which are  responsible for more than one  
self-consistent value for adsorbate charge.[21,31,32].
The  calculations show  that all the three roots of $<n_{s\sigma}>$  survive for all
coverages if 
the desolvation of the adsorbate with the increasing coverage is disallowed. In fact
it is this  progressive desolvation which leads to the transition from multiple to a
single charge state of copper.
In the present formalism the desolvation effect has been modeled through
the  scaling factor $ (1 - \theta^2)$.
But the occurrence of this  transition does not depend on this  specific 
form of the scaling factor. Any factor $ f(\theta)$ will do ,
provided $ f(\theta) \rightarrow 0 $ when $\theta \rightarrow 1 $, and $f(\theta)
\rightarrow 1 $ when $\theta \rightarrow 0$. These limits imply that in the former case, the 
solvation energy of the adsorbate is $\alpha \Delta G_s $ and the same is zero when
a metallic adsorbate forms a monolayer. We have, however, taken $ f(\theta)$ as
$(1 - \theta^2)$ since this implies that the decrease in the solvation energy of
the adsorbate depends on the probability of finding another adsorbate in its vicinity.

The  presence of multiple charge state for the copper s orbital, and the subsequent
transition to a single charge state beyond a critical coverage,
also get reflected in the binding energy
$\Delta E_s$ of the copper s orbital in the electrochemical environment. The
change in $\Delta E_s$ with respect to coverage is substantial. But more importantly,
$\Delta E_s$ is positive for all coverage regime. Therefore, the consideration of
s orbital alone would  not lead to the electrochemisorption of copper. This contrasts
with the adsorption at the metal-vacuum interface wherein the s orbital bonding
provides the major contribution to the chemisorption energy. 
The difference in the
s orbital bonding nature at the metal-vacuum and electrode-electrolyte interface is a consequence of the fact that
in the former case, a far separated copper adsorbate is in a neutral state, whereas in
the latter case, s orbital is unoccupied in a  solvated copper ion residing deep
inside the bulk electrolyte. The desolvation of adsorbate upon chemisorption and
energy difference between a neutral copper  atom and a solvated $Cu^+$ ion leads to
positive energy value for s orbital bonding. 
The binding energy contribution $\Delta E_d$
for copper d orbital does not vary much with coverage and it is  sufficiently negative
($\approx-4.4~ eV$). The negative chemisorption energy for the d orbital results 
from the large difference between the energies of the  copper d orbital and gold Fermi
level. Upon electrochemisorption, an unoccupied copper  d orbital gets almost completely
filled and this leads to substantial decrease in the  energy.

The total binding energy $\Delta E$ is a negative quantity, thus making
electrosorption of copper on gold electrode feasible (cf. Figure 6), whatever may be
the  charge state of the copper adsorbate. But as we have noted earlier, configurations
having more ionic character  are more stable, with the positive charge state being most
stable.
The calculations have also shown that $\Delta E$ at the electrode-electrolyte interface
is more negative than the $\Delta E$ at the metal-vacuum interface.
Finally, we have also verified that a decrease in $\alpha$, the degree of solvation
for  an interfacial adsorbate, reduces the  binding energy.

The analysis presented here  unambiguously supports the Tadjeddine et al.
in situ experimental results  showing a  positive charge copper adatom when the coverage
is large [6]. In this domain, only one self-consistent value of $<n_{s\sigma}>$ exists.
For the low coverages, the theory predicts that apart from ionic configurations,  copper
adsorbate can also exist in a relatively neutral state.
But the  energetic considerations make this state to be unstable.
In fact when $v_s \rightarrow 0 $, the stability of the various possible charge states
of copper decreases  in the order $ Cu^{+} >  Cu^{-} >  Cu^o $. The subsequent increase
in $v_s$ does not alter this trend; it only lowers the  respective energy values.
So it seems that Tadjeddine et al., experimental observation 
of a positive charge copper adsorbate at  higher coverage should be valid even for small
$\theta$ (including the case of a single adsorbate). In such a case, the  
net charge on copper would decrease in a smooth way with the increasing 
coverage (curve a, Figure 5).

On the other hand, electrosorption valence data suggest that copper is neutral when the 
coverage is low. But this still requires a justification why the copper adsorbate on 
gold electrode should exist in a metastable configuration at low coverage.  
Notwithstanding the above  energetic considerations, if the copper adsorbate is indeed 
neutral for small $\theta$, the present formalism provides a method to verify 
this situation. The analysis carried out so far regarding the
charge variation with coverage shows that the relatively neutral copper adatom 
first becomes  more negative with increasing coverage. But beyond a critical coverage,
it acquires a positive charge in a discontinuous fashion (cf. Figure 5). So in
case copper adsorbate exists in two different charge configurations in the low
and higher coverage regime, the switch over between them
is not  smooth. A sharp valence transition for  the copper 
adatom in the intermediate coverage region thus signifies the  presence of a neutral  
copper when coverage is low. We also note that decrease in $v_s$ shifts this critical
coverage to higher values.

\subsection{ Wide band limit}

~~~~We now provide a brief discussion on the applicability of the wide-band limit
to copper adsorption on gold. We start with the lone adsorbate case (cf. eq 6.16a)
Following refs 17 and 18, we take $\Delta_s$ as 1.0 eV. For the chemisorption
at the M-V interface, only one self-consistent value of $< n_{s\sigma}>$ equal to
0.49 is obtained. That the $< n_{s\sigma}>$ lies near to 0.5 can be ascertained
by comparing the copper s orbital energy with $\Delta_s$. Taking the energy zero 
at the Fermi level, the $\epsilon_{s \sigma}$ is 0.04 eV when $<n_{s\sigma}>$ is
0.5. As this value is much lower than $\Delta_s$, the self-consistent expression

$$
< n_{i\sigma}> = (1/ \pi) cot^{-1}(\epsilon_{i\sigma} / \Delta_i) ~;~~~ \{i =  s, d \}
$$
$$
\epsilon_{i\sigma} = \epsilon_{a}^o - z_i\epsilon_{im} + (U_i + 2\epsilon_{im})<n_
                     {i \bar{\sigma}}> ~;~ ~~\{ z_i = 1(3)~ for~s(d)-orbital \} 
  \eqno 8.3
$$

\noindent leads to $< n_{s\sigma}>~ \approx $  0.5 in the non-magnetic limit. 

The lower edge of conduction band $\epsilon_{\Gamma}$ for gold lies at 8.10 eV below 
the Fermi level [22]. With this cutoff,
the s orbital binding energy $\Delta E_s$ when the charge is 0.49 is - 0.74 eV only.
This binding energy at M-V interface is obtained by using the relation   

$$ \begin{array}{lcl}
\Delta E_i & = &  \sum_{\sigma} \epsilon_{i\sigma} <n_{i \sigma}> - (U_i + 2\epsilon_{im})
       <n_{i\sigma}><n_{i \bar{\sigma}}> + \sum_{\sigma} (\Delta_i/2 \pi)
     ~  \ln  \frac {\displaystyle \epsilon_{i \sigma}^2 + \Delta_i^2}
               {\displaystyle ( \epsilon_{\Gamma} - \epsilon_{i\sigma})^2 + \Delta_i^2}\\
\\
& + &  x_i^2 \epsilon_{im} - \sum_{\sigma} \epsilon^o_{i\sigma}< n_{i\sigma}>^o - 
U_i <n_{i\sigma}>^o <n_{i \bar{\sigma}}>^o  ~ ;~~~ \{x_i = 1(2)~ for~s(d)- orbital\} \\
\end{array}    \eqno 8.4 
$$

The above expressions for the average charge and binding  energy for a single adsorbate
at M-V interface follows from eqs 6.16a, 6.18, 5.15, and 7.4.

For the copper d orbital, the self-consistent $<n_{d \sigma}>$ at the M-V interface
is 0.96 when $\Delta_d$ = 1.0 eV. The corresponding binding  energy $\Delta E_d$
is +1.98 eV. If $\Delta_d$ is taken as 2.0 eV, the above two quantities are
0.925 and +3.07 eV respectively. The reason behind getting such positive values
for $\Delta E_d$ is that the argument of the ln function in  eq 8.4 is greater than
unity when the cutoff is taken at the bottom of conduction band. The binding 
energy value is also sensitive to the lower cutoff limit. For example, if the cutoff
is taken arbitrarily  at  -15.0 eV, the $\Delta E_d$ gets reduced from +3.07
to +1.48 eV. This strong dependence of binding energy on cutoff parameter is
also true for the copper s orbital. With lower cutoff at -15.0 eV, $\Delta E_s$
corresponding to $\Delta_s$ = 1.0 eV is -1.13 eV. Finally in all the cases, the total
binding energy remains positive.

Thus we find that the wide-band approximation predicts no  binding 
even for the single adsorbate case at the  M-V interface, and the copper adsorbate
exists in almost neutral state. 

Application of this formalism  to a lone copper adsorbate  at gold electrode
again gives rise to  three self-consistent solutions for $<n_{s\sigma}>$ at 0.085
($\Delta E_s$ = 1.25 eV), 0.51 ($\Delta E_s$ = 2.21 eV), and 0.91 ($\Delta E_s$ = 2.00 eV)
when $\Delta_s$ = 1.0 eV and the lower cutoff is placed at the $\epsilon_{\Gamma}$
in the energy calculations. The corresponding $<n_{d\sigma}>$ is 0.96 
($\Delta E_d$ = -2.14 eV) for  $\Delta_d$ = 1.0 eV. 

The above calculation again shows that as  $\theta \rightarrow 0$ , the  $Cu^{+0.91}$
is the most stable configuration, followed by  $Cu^{-0.74}$.
The nearly neutral charge state of copper ($Cu^{+0.06}$) is the least stable one.  
The respective total binding energies corresponding to these three charge configurations
are -0.89, -0.14, and 0.07 eV.

Calculations based on eqs 4.3-4.4, and 6.16 show that 
the two roots of $<n_{s\sigma}>$ again merge at $\theta $ = 0.63. Beyond this coverage,
the $<n_{s\sigma}>$
is  single valued, and it equals  0.23, 0.39, 0.474, 0.486, and 0.49 when
$\theta$ is 0.64, 0.7, 0.8, 0.9, and 1.0 respectively. Thus wide-band approximation
leads to a relatively neutral configuration of copper in the high-coverage regime.

The above analysis shows that the gross features like multiple roots for 
$<n_{s\sigma}>$ at the E-E interface, relative stability of these solutions, and existence of
a single solution  beyond a critical coverage  can be obtained through the wide-band
limit based approach. But it gives poor result for binding energy. Besides, it predicts
the existence of almost neutral copper adsorbate for large coverages which is 
contrary to the observations of Tadjeddine et al. [6].

\subsection{Adsorbate's Density of States}

~~~~The present calculations have shown that the average occupancy of copper d orbital
remains constant with respect to change in the coverage, whereas $<n_{s\sigma}>$
exhibits steep variation. The underlying reason for this  feature can be
understood through the respective density of states. The copper   d  orbital density 
of states for $\theta$ = 0.1 in the electrochemical environment is plotted in Figure 7.
The two peaks in the density of states, lying below the Fermi level (at 5.20 eV above
the energy zero which is taken at the gold d band center) are to be  
ascribed to the bonding and antibonding states, broadened due to their interactions
with the substrate band. Only the tail portion of the density of states lies near
the Fermi level. A change in the coverage, though inducing a slight shift in the
peaks positions and reducing their heights, does not alter the density of states
near the Fermi level. Thus the d orbital average occupancy, which is the
integrated density of states up to Fermi level, remains constant with the varying
coverage. When $\theta$ = 0.10, we get three  density of states for the copper s orbital
corresponding to the three self-consistent values of $< n_{s \sigma}>$ at E-E
interface . The density of states (dos) for $<n_{s \sigma}>$ = 0.04 and 0.99 lie much above
and below the Fermi level, respectively (Figure 8). The former has a single narrow peak
and  the latter is broad and exhibits a diffuse satellite peak.
The dos  $\rho_{a}$ corresponding to $<n_{s \sigma}>$  = 0.59 is peaked around the Fermi level
and has a small satellite peak. As the coverage is increased, peaks in the dos for the 
least and largest $<n_{s\sigma}>$ move toward the Fermi level  whereas the dos peak
corresponding to middle root shifts below the $\epsilon_f$ (Figure 9).
This explains the ensuing variation in the adsorbate charge with coverage.
The ionic configurations tend to reduce their net charge,
whereas the neutral state becomes more negative with increasing $\theta$ (cf. Figure 5).
Note that the increase in $\theta$ leads to a broader density of states along with 
reduced peak height. Whenever a dos peak lies in the vicinity of $\epsilon_f$,
any change in its position or shape, resulting either from the increase in coverage,
or through variations in the other system parameters, has a pronounced effect on
the magnitude of the corresponding charge.
This explains the large variation in the net charge  Q 
in different coverage domains (branches a, b in Figure 5).

The adsorbate density of states are primarily considered here to explain the trends in
the adsorbate average occupation probability, and it is also needed for the binding energy
calculation. But $\rho_{a}$ also plays a central role in the charge transfer processes
involving metal-adsorbate complex, be it radiative, nonradiative, or related to STM studies.
In this context, earlier results for the adsorbate dos  by Schmickler, Kuznetsov, and Ulstrup 
for electron transfer via adspecies [31,33-35] and the 
more recent work by Schmickler for  STM studies [36,37] of redox couple and in situ
electron tunneling through thin water layer at electrode are notable. Recently
we have obtained a closed form expression for $~\rho_{a}~$ within the semiclassical
approach for solvent and wide-band approximation. The density of states thus derived
admits Lorentzian, Gaussian, and delta function structure in appropriate limits [38]. 
The voltage spectroscopy of adsorbate ions, preferably through STM measurements,
can give information about the in situ adsorbate density of states [39].
Though this approach has been used extensively to obtain $\rho_{a}$ at the M-V interface [40], 
to our knowledge no such experimental result in the context of E-E interface is available.
We shall report in future papers how  the various electron transfer processes
noted above are affected by the coverage dependence of the adsorbate density of states.

\section{Summary and Conclusions}

~~~~In order to understand the electronic structure of a chemisorbed layer in an
electrochemical environment, an appropriate model Hamiltonian based formalism is
constructed. Starting with the lone adsorbate case $(\theta \rightarrow 0)$, the
formalism is valid up to monolayer regime $(\theta =1)$. The adsorbates are  assumed to
occupy the on-top positions over the substrate atoms and the adsorbate layer is
commensurate with the underlying electrode surface lattice. At coverages less than
unity, the adsorbates are distributed randomly over various adsorption sites,
and give rise to a 2D electron band in the monolayer regime. The randomness in the 
system is treated self-consistently through the CPA approach. The adsorbate self-energies
arising due to  its hybridization with the substrate states are evaluated using 
appropriate density of states for the substrate band. This allows us to go beyond the 
wide-band approximation, removes the divergence associated with the binding energy
calculation in the wide-band limit, and leads to non-Lorentzian adsorbate density of states.

The formalism is applied to a chemisorbed layer of copper  on gold electrode. 
The following are our main  results:

\noindent 1. \ Double layer  studies have suggested that at low coverages, 
copper electrochemisorbed on gold electrode exists in a neutral state. The in situ
X-ray absorption spectroscopy indicates that in the monolayer regime, copper adion is
positively charged. Present  theoretical analysis supports the latter. When the
coverage is low, three self-consistent charge states for copper are possible. Though
one of them is relatively neutral, binding energy calculations make it to be the least stable.
The positive charge configuration of copper is most stable at all  coverages.

\noindent 2. \ The transition from multiple to unique charge configuration for copper
is caused due to the progressive desolvation of copper adion with the increasing coverage.
In spite of the energetic considerations, 
if the  copper adsorbate is indeed 
neutral in the low and positively charged in the higher 
coverage domains, the switchover between these two configurations
is not  smooth. It occurs through a  sharp valence transition in the
copper adatom in the intermediate-coverage region.

\noindent 3. \ The coverage variation has a little effect on copper d-orbital
occupancy. All the different charge configurations of the copper adsorbate result due to 
changes in its s orbital occupancy with varying coverage.

\noindent 4. \ Density of states for the d orbital of the chemisorbed copper  exhibits a
two-peak structure. Both the peaks lie well below the Fermi level of gold. Thus, a shift 
in the peak positions, either because of the coverage variation or due to change in other
parameters, would have minimal effect on the observables pertaining to the copper d orbital.
On the other hand, the major peak corresponding to at least one of the density of states 
for s orbital always lies near the Fermi level. Consequently, the s orbital properties are
more sensitive to the coverage and other parametric variations.

\noindent 5. \  Calculations have shown that for the copper chemisorption on gold at the 
metal-vacuum interface, it is the s orbital bonding which contributes maximally towards
the binding energy. But in the electrochemical environment, the binding energy for
the s orbital is positive. The binding is possible only when the d orbital energy
contributions are also taken into account in the energy calculations.

\noindent 6. \ The wide-band limit based approach has been generalized to study the electronic
properties of a random chemisorbed layer. The subsequent analysis shows that it gives poor  
results for binding energy. Besides, it predicts 
the existence of almost neutral copper adsorbate for large coverages which is 
contrary to the experimental observations of Tadjeddine et al. [6]. 

We have presently treated the randomness in the system through coherent potential
approximation, which is an effective mean-field theory. To take into account the
possible short-range order in the system, one  needs  to go beyond the CPA approach.

The copper adsorption in the presence of anions and the electron transfer through 
adsorbed layer represent possible  extensions of the present work. Our results
on these important problems will be reported in future communications.
  
\vskip 0.5 cm

\noindent {\bf Acknowledgment} I thank 
Prof. S.K.Rangarajan for providing valuable insights and  many  
enlightening discussions. I am also indebted to  all the reviewers 
for their many constructive comments and criticism which have definitely helped me
in improving the paper.

\newpage

\pagestyle{empty}
\centerline{\bf TABLE \ 1}

\vskip .5cm


{\tiny{\bf
\begin{center}
\begin{tabular}{|c|c|c|c|c|c|c|c|} \hline 
& & & & & & &  \\
$\phi$ & $\Delta_d$ & $\epsilon_f$ & $E_s$ & $E_d$ & \begin{tabular}{c}$\Delta G_s$ \\ 
$(C^+_u)$\end{tabular} & $\epsilon_{im}$   & $U_s$  
\\ 
\hline
& & & & & & & \\
4.80 & 5.60 & 5.20$^\dagger$ & 2.28$^\dagger$ & -10.14$^\dagger$ & 5.92 & -1.76 & 5.92  
 \\
& & & & & & & \\ \hline
\end{tabular}
\end{center}
}}

{\tiny{\bf
\begin{center}
\begin{tabular}{|c|c|c|c|c|c|c|c|} \hline 
& & & & & & &  \\
$U_d$ & 
$\Delta_{||}$ & $\Delta_{\perp}$ & $v_{sd}$ & $v_{dd}$ & \begin{tabular}{c}$\alpha$  \\ $~~~~~$\end{tabular} & $\mu_s$ & $\mu_d$ \\
\hline
& & & & & & &  \\
5.45~ & 
1.40~ & 1.40~ & ~2.68~ & 4.85~ & 0.50~ & 0.7~5 & 0.60~  \\
& & & & & & & \\ \hline
\end{tabular}
\end{center}
}}

\vskip .3cm

\noindent $^\dagger$ These energies are relative to the d band centre of Au. $\alpha,~ \mu_s$ and $\mu_d$ are 
dimensionless quantities.  All other quantities are in the units of eV. $\alpha$ and
$\Delta G_s$ are zero for chemisorption at a metal-vacuum interface. The $\epsilon_{im}$
is for the M-V interface.

\newpage


\noindent {\bf Figure Captions }

\begin{description}
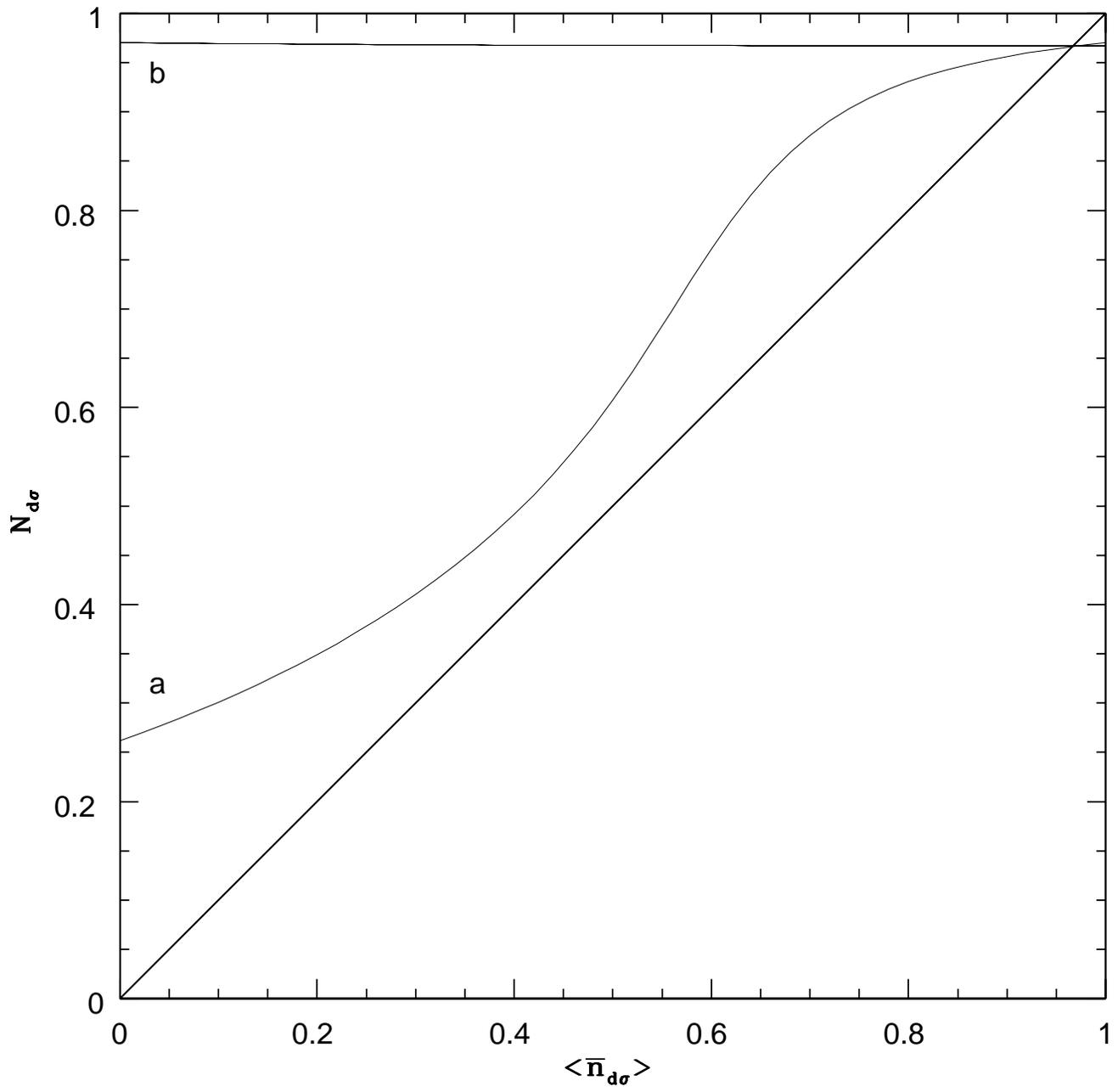
\item[{FIG.1}] Self-consistent solution to d orbital occupancy of Cu
chemisorbed on $Au(111)$ electrode surface at electrochemical interface: (a) $\theta$ = 0.10,
$<n_{d\sigma}>$ = 0.965; (b) $\theta$ = 0.90, $<n_{d\sigma}>$ = 0.967.
Refer to  TABLE 1 for parameters.
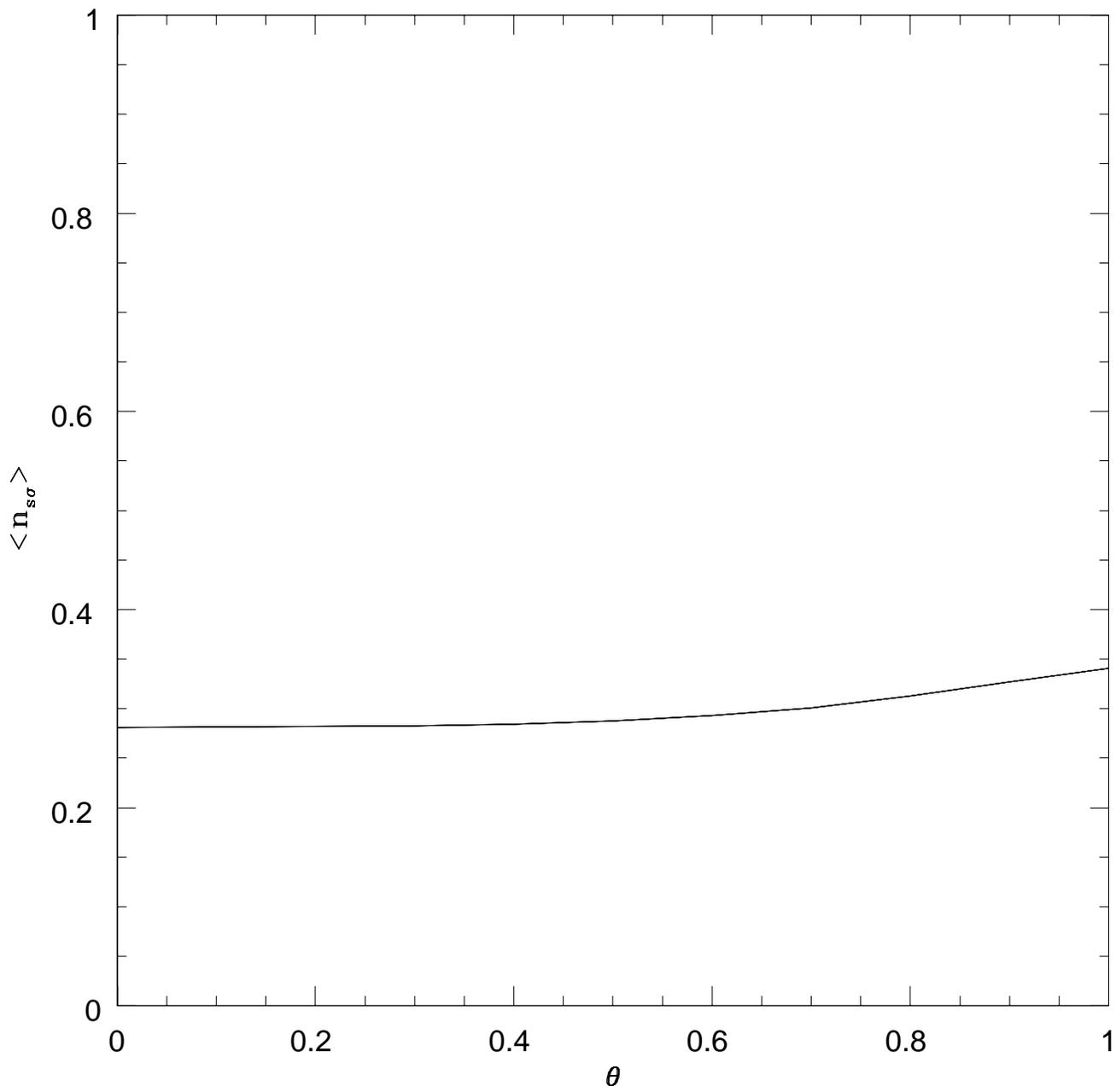
\item[{FIG.2}] Coverage dependence of average s orbital occupancy 
$<n_{s\sigma}>$ 
for Cu chemisorbed on Au(111) surface at the metal-vacuum interface.   
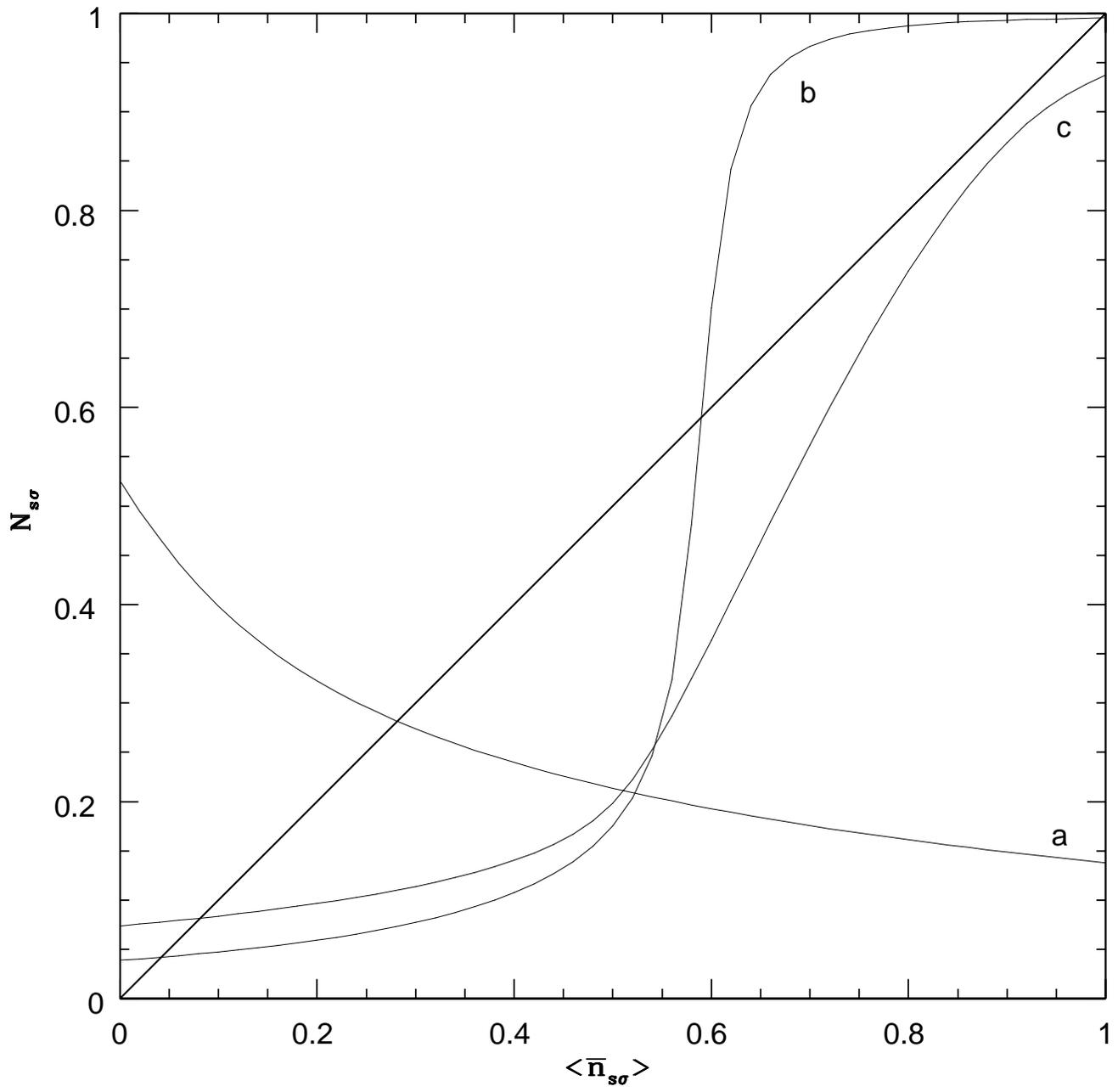
\item[{FIG.3}] Self-consistent solutions to s orbital occupancy of Cu
chemisorbed on Au(111): (a) $\theta$ = 0.10, $<n_{s\sigma}>$ = 0.285 (M-V
interface, $v_s$ = 2.70 eV ); (b) $\theta$ = 0.10, $<n_{s\sigma}>$ = 0.04,
0.59 and 0.99 (E-E interface, $v_s$ = 2.0 eV); (c) $\theta$ = 0.67, 
$<n_{s\sigma}>$
= 0.082 (E-E interface, $v_s$ = 2.0 eV).
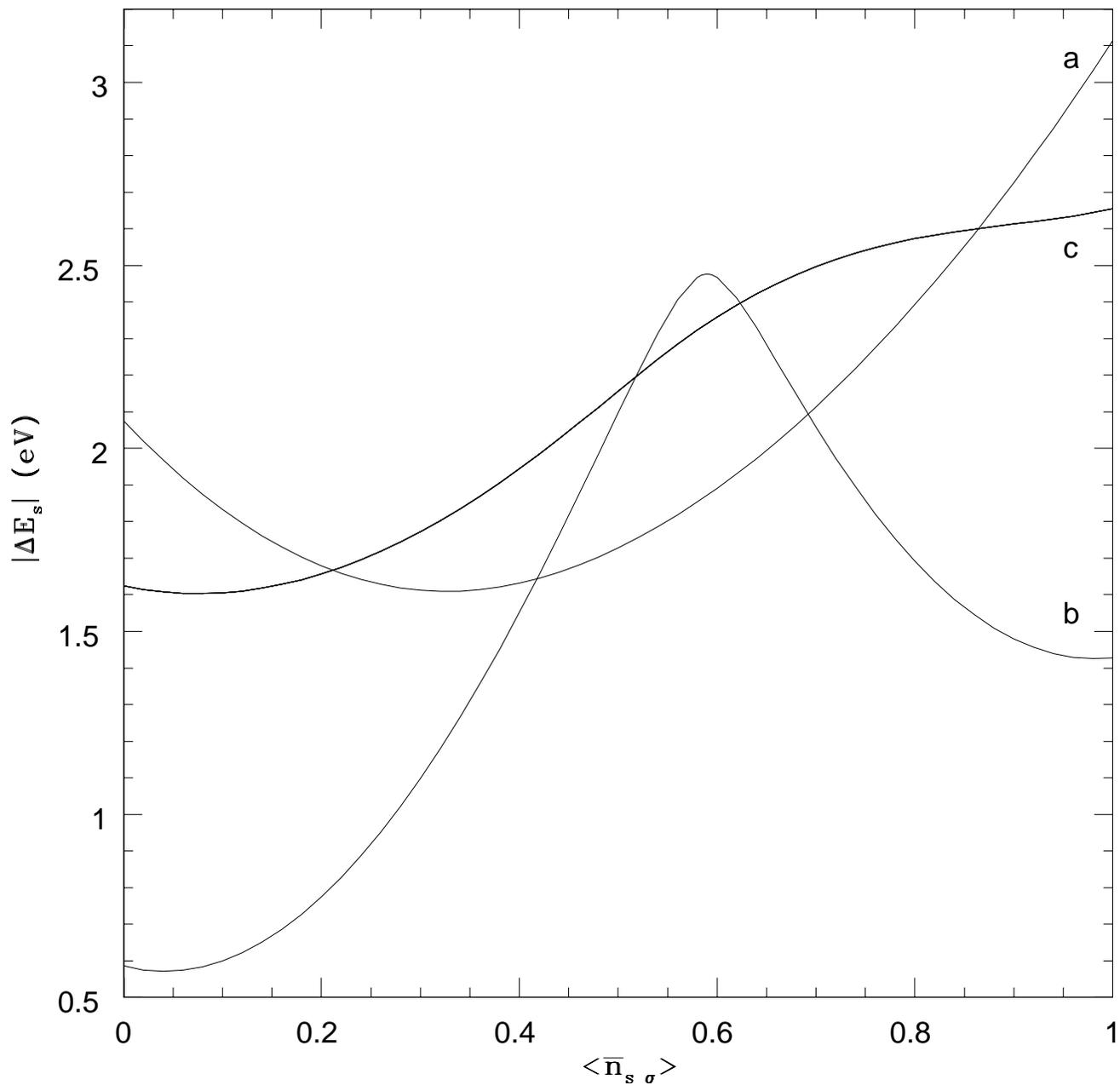
\item[{FIG.4}]Extremum in the binding energy $\Delta E_s$ at the self-consistent values  of
average electronic charge on copper s orbital. (a) $\theta$ = 0.90, 
$<n_{s\sigma}>$
= 0.327 (M-V interface, $v_s$ = 2.70 eV); the parameters for (b) and (c) are same 
as for  curves b and c in the Figure 3 .
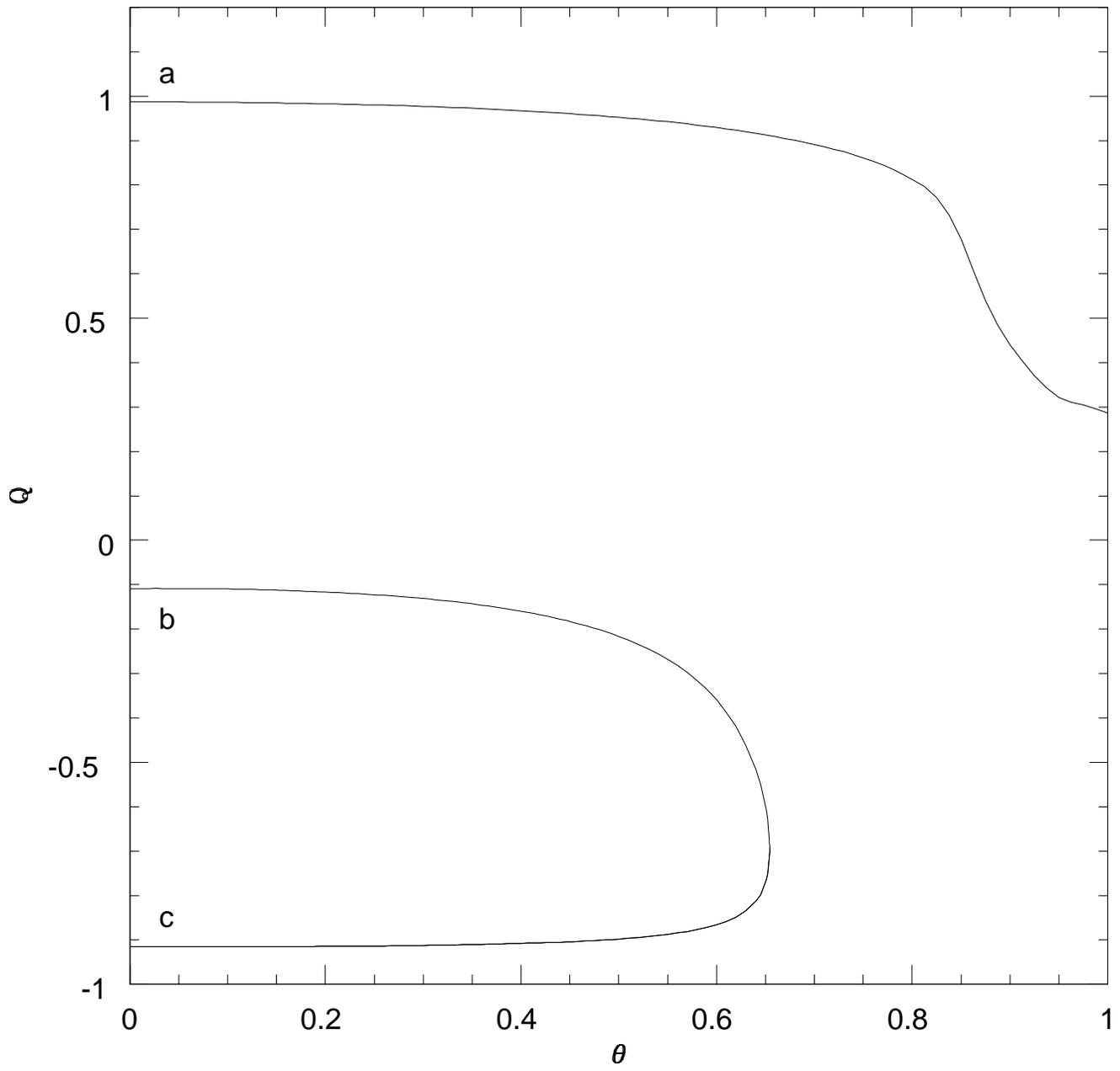
\item[{FIG.5}] Total charge Q on Cu electrochemisorbed on Au(111) electrode:
coverage dependence. The branches (a), (b), and (c) correspond to multiple 
self-consistent values of $<n_{s\sigma}>$ ($v_s$ = 2.0 eV). (a) and (b) merge
together at $\theta$ = 0.65 where the common value of Q is -0.69.
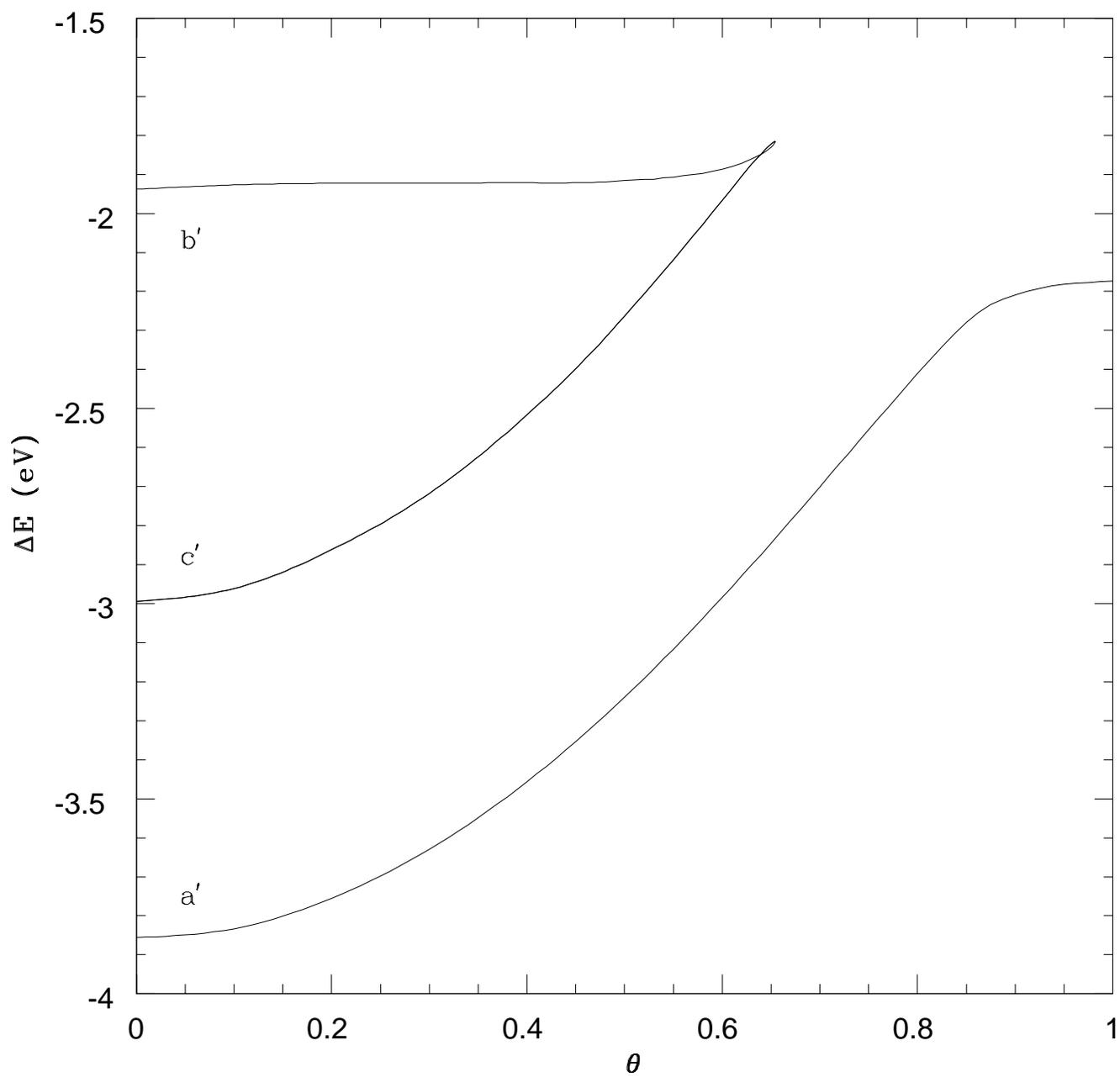
\item[{FIG.6}] Coverage dependence of the total binding energy $\Delta E$
for Cu on Au(111)
at electrochemical interface. The curves $a^{\prime}, b^{\prime}$,  and $(c^{\prime})$
here depict the respective
binding energies corresponding to the branches a, b, and c in Figure 5 for the
total charge on copper.
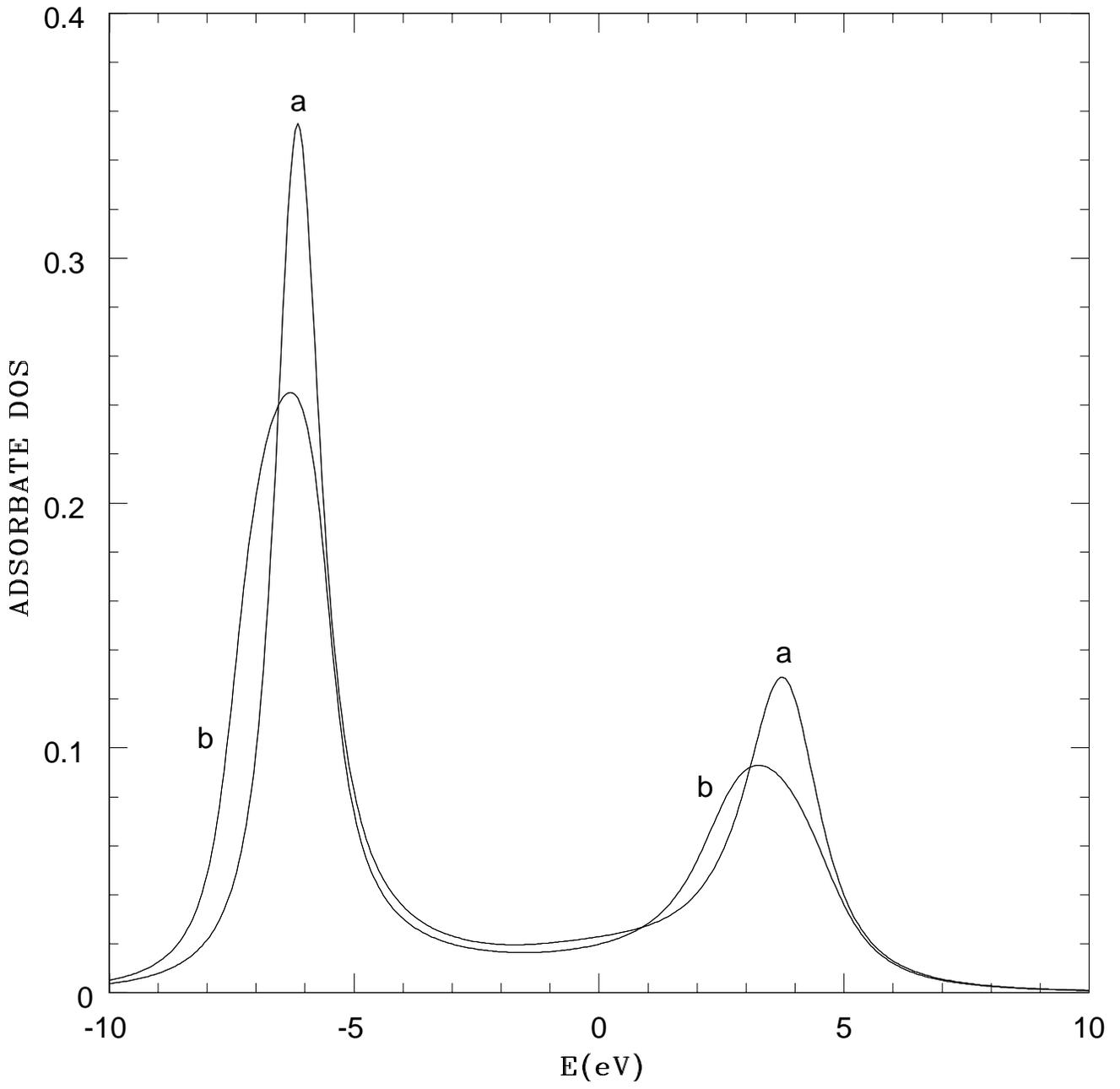
\item[{FIG.7}] d orbital density of states for Cu electrochemisorbed on
Au(111) electrode: (a) $\theta$ = 0.10, $<n_{d\sigma}>$ = 0.965; (b) $\theta$
= 0.90, $<n_{d\sigma}>$ = 0.967. Au d band centre is at 0.0 eV. $\epsilon_f$ = 5.20 eV,
vacuum level at 10.00 eV.
 
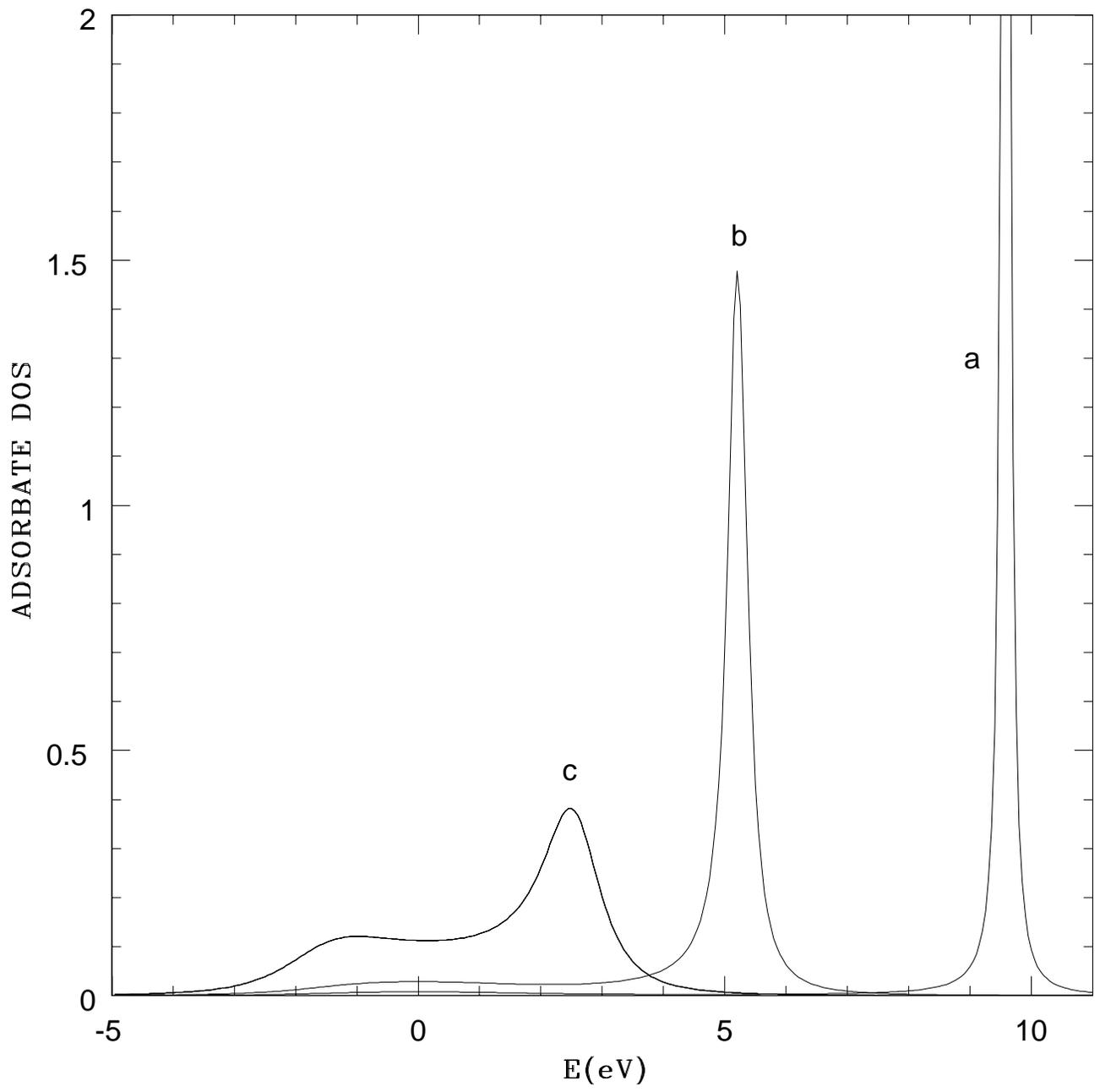
\item[{FIG.8}] s orbital density of states for Cu electrochemisorbed on
Au(111) electrode.  $\theta$ = 0.10, $v_s$ = 2.0 eV. (a) $<n_{s\sigma}>$ = 0.04,
the height of dos peak is 4.01;
(b) $<n_{s\sigma}>$ = 0.59; (c) $<n_{s\sigma}>$ = 0.99.
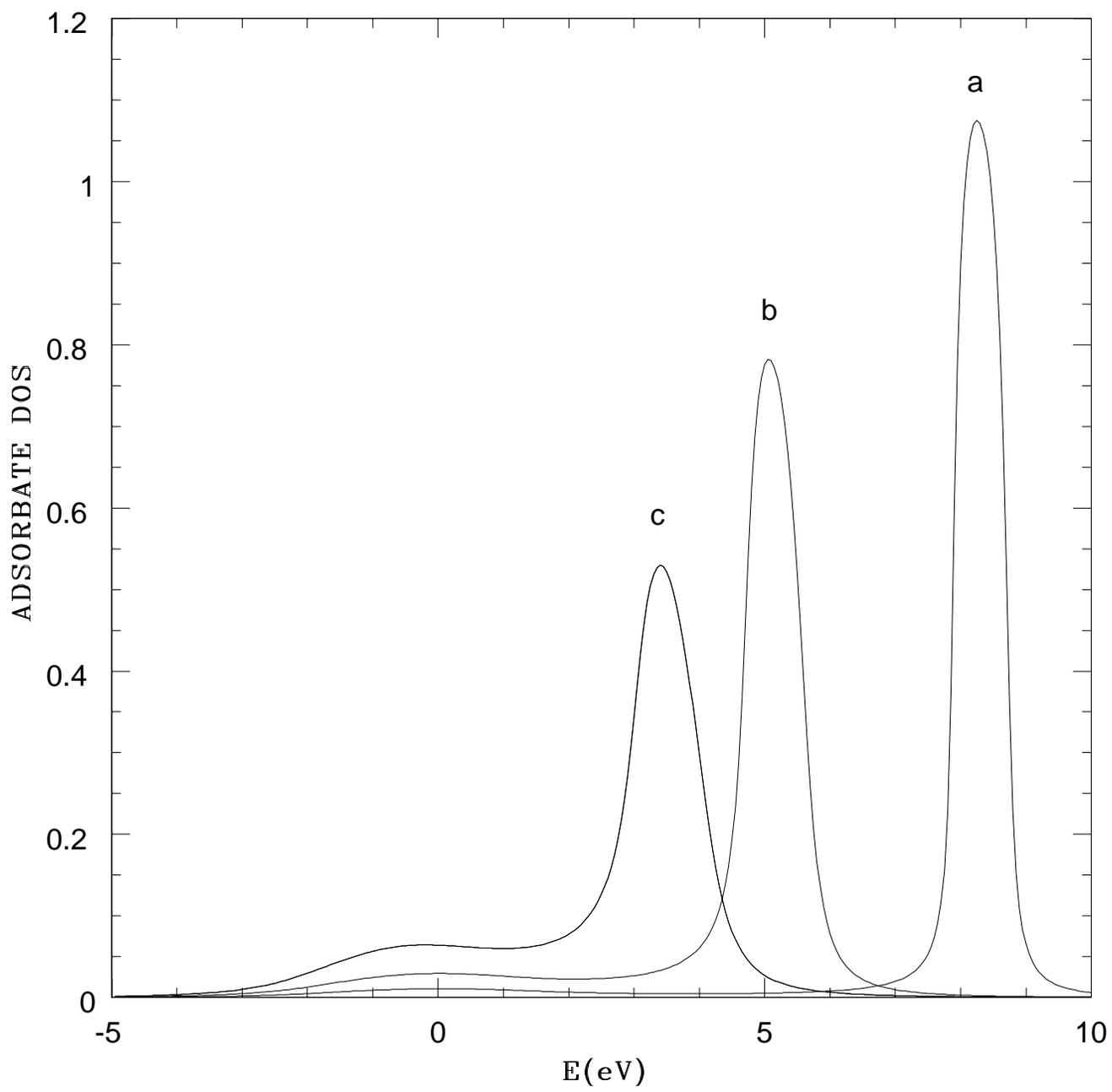
\item[{FIG.9}] s orbital density of states for $Cu$ electrochemisorbed on
$Au(111)$ electrode.  $\theta$ = 0.50, $v_s$ = 2.0 eV. (a) $<n_{s\sigma}>$ = 0.06;
(b) $<n_{s\sigma}>$ = 0.64; (c) $<n_{s\sigma}>$ = 0.98.
\end{description}

\end{document}